# Photonic Networks-on-Chip Employing Multilevel Signaling: A Cross-Layer Comparative Study

Venkata Sai Praneeth Karempudi

Department of Electrical and Computer Engineering, University of Kentucky, kvspraneeth@uky.edu

Febin Sunny

Department of Electrical and Computer Engineering, Colorado State University, febin.sunny@colostate.edu

Ishan G Thakkar

Department of Electrical and Computer Engineering, University of Kentucky, igthakkar@uky.edu

Sai Vineel Reddy Chittamuru

Micron Technology, Inc., Austin, TX, USA, vineelreddys@gmail.com

Mahdi Nikdast

Department of Electrical and Computer Engineering, Colorado State University, Mahdi.nikdast@coloradostate.edu

Sudeep Pasricha

Department of Electrical and Computer Engineering, Colorado State University, sudeep@colostate.edu

Photonic network-on-chip (PNoC) architectures employ photonic links with dense wavelength-division multiplexing (DWDM) to enable high throughput on-chip transfers. Unfortunately, increasing the DWDM degree (i.e., using a larger number of wavelengths) to achieve higher aggregated datarate in photonic links, and hence higher throughput in PNoCs, requires sophisticated and costly laser sources along with extra photonic hardware. This extra hardware can introduce undesired noise to the photonic link and increase the bit-error-rate (BER), power, and area consumption of PNoCs. To mitigate these issues, the use of 4-pulse amplitude modulation (4-PAM) signaling, instead of the conventional on-off keying (OOK) signaling, can halve the wavelength signals utilized in photonic links for achieving the target aggregate datarate while reducing the overhead of crosstalk noise, BER, and photonic hardware. There are various designs of 4-PAM modulators reported in the literature. For example, the signal superposition (SS), electrical digital-to-analog converter (EDAC), and optical digital-to-analog converter (ODAC) based designs of 4-PAM modulators have been reports. However, it is yet to be explored how these SS, EDAC, and ODAC based 4-PAM modulators can be utilized to design DWDM-based photonic links and PNoC architectures. In this paper, we provide a systematic analysis of the SS, EDAC, and ODAC types of 4-PAM modulators from prior work with regards to their applicability and utilization overheads. We then present a heuristic-based search method to employ these 4-PAM modulators for designing DWDM-based SS, EDAC, and ODAC types of 4-PAM photonic links with two different design goals: (i) to attain the desired BER of $10^{-9}$ at the expense of higher optical power and lower aggregate datarate; (ii) to attain maximum aggregate datarate with the desired BER of $10^{-9}$ at the expense of longer packet transfer latency. We then employ our designed 4-PAM SS, 4-PAM EDAC, 4-PAM ODAC, and conventional OOK modulators based photonic links to constitute corresponding variants of the well-known CLOS and SWIFT PNoC architectures. We eventually compare our designed SS, EDAC, and ODAC based variants of 4-PAM links and PNoCs with the conventional OOK links and PNoCs in terms of performance and energy-efficiency, in the presence of inter-channel crosstalk. From our link-level and PNoC-level evaluation, we have observed that the 4-PAM EDAC based variants of photonic links and PNoCs exhibit better performance and energy-efficiency compared to the OOK, 4-PAM SS, and 4-PAM ODAC based links and PNoCs.

**ACM Reference Format:**

# 1 Introduction

As the core count in contemporary manycore processing chips increases, the conventional on-chip communication fabrics, i.e., electrical networks-on-chip (ENoCs), experience higher power dissipation and degraded performance. As a potential solution to these shortcomings, ENoCs have been projected to be replaced by emerging photonic network-on-chip (PNoC) fabrics. This is because the recent advancements in silicon photonics have enabled PNoCs to offer several advantages over ENoCs, such as higher bandwidth density, distance-independent datarate, and smaller bandwidth-dependent energy.

Typical PNoC architectures (e.g., [1]-[4], [8],[59]) and processor-to-DRAM photonic interconnects (e.g., [25], [26]) utilize several photonic devices such as multi-wavelength lasers, waveguides, splitters and couplers, along with microring resonators (MRs) as modulators, detectors and switches. A broadband laser source generates light of multiple wavelengths (λs), with each wavelength (λ) serving as a data signal carrier. Simultaneous traversal of multiple optical signals across a single photonic waveguide is possible using dense wavelength-division multiplexing (DWDM), which enables parallel data transfers across the photonic waveguide. For instance, a DWDM of 16 λs in the photonic waveguide can transfer 16 data bits in parallel. At the source node, multiple MRs modulate multiple electronic data signals on the utilized multiplexed λs (data-modulation phase). In almost all PNoC architectures in literature, modulator MRs utilize on-off keying (OOK) modulation, wherein the high and low intensities of λs in the waveguide are used to represent, respectively, logic '1' and '0'. Similarly, at the destination node, multiple MRs equipped with photodetectors are used to filter and detect λ-modulated data signals from the waveguide (data-detection phase) and convert them back to proportional electrical signals. In general, using a large number of multiplexed λs enables high-throughput parallel data transfers in PNoCs, hence boosting the bandwidth in such networks.

Leveraging a large number of multiplexed λs, and thus the resultant high throughput, has been pivotal in PNoC architectures for efficiently amortizing their high non-data-dependent power consumption that includes the laser power and MR tuning power. However, a number of challenges related to area [8], cost [32], reliability [11], and energy-efficiency [13][60] still need to be overcome for efficient implementation of PNoCs that utilize a large number of multiplexed λs (typically 32 or more multiplexed λs per waveguide [35], [53]). First, generating a large number of multiplexed λs requires a comb laser source, the ineffectiveness, complexity, and cost of which increase with the number of generated λs [5]. Second, utilizing a larger number of multiplexed λs to achieve higher-throughput data transfers in a PNoC results in higher area and power overheads. A large number of multiplexed λs require larger network flit size as well as more electrical and photonic hardware such as modulator and detector MRs and their drivers. A larger network flit size can also result in larger sized electronic buffers in the network gateway interfaces, which can result in significantly higher area and power overheads. Similarly, larger number of MRs and drivers also incur greater photonic area and MR tuning power overheads. Last, the use of a larger number of multiplexed λs can decrease the viable gap between two successive optical signals, which in turn will increase the inter-channel crosstalk noise in PNoCs, increasing the bit-error rate (BER) of communication [6], [9]. As a result of the combined impact of these factors, the use of larger number of multiplexed λs in PNoCs leads to tradeoffs among the achievable throughput, BER, and energy-efficiency.

To mitigate the adverse impacts of these tradeoffs, multi-level optical signaling has been introduced in prior works ([4],[7],[21]-[22],[36],[61],[76],[77]). For example, in [4] and [7], Kao et al. proposed a multilevel optical signaling format four-pulse amplitude modulation (4-PAM) to achieve higher-throughput and energy-efficient data communication in PNoCs. The 4-PAM optical signaling format doubles the datarate by compressing two bits in one symbol carried out by four levels of optical intensity. In the literature, three different MR-based designs of optical 4-PAM modulators have been proposed. In [36] and [37], two cascaded on-off keying (OOK) modulators are utilized to superimpose two OOK optical signals of the same λ with 2:1 power ratio to create a 4-PAM λ-signal. But this signal superposition based 4-PAM method (referred to as 4-PAM-SS henceforth) incurs substantially high power, photonic area, and reliability overheads at the link-level, as discussed in Section 3. Roshan-Zamir et al. in [21] demonstrated a single-MR 4-PAM modulator that takes an electrical 4-PAM signal, generated using a segmented pulsed-cascode amplifier based electrical DAC (EDAC), as input and then converts it into an optical 4-PAM signal. But this EDAC-based conversion method (referred to as 4-PAM-EDAC henceforth) can incur significant power consumption and area overheads due to the required EDACs (Section 3). In contrast, Moazeni et al. [22] utilized an optical DAC (ODAC) modulator (referred to as 4-PAM-ODAC henceforth) that directly converts two input electrical OOK signals into a 4-PAM optical signal, thereby eliminating the use of EDAC and its overheads. Thus, it is well established how various MR-based modulators can be utilized to generate 4-PAM optical signals. But what is still unknown is how different 4-PAM modulators can be utilized to design DWDM-based photonic links and PNoC architectures. Moreover, the impacts of various 4-PAM modulators on the overall energy, reliability, and performance behavior of the designed links and PNoC architectures also remain unexplored.

In this paper, we present a comparative study and a heuristic-based search method for designing DWDM-based on-chip photonic links using different types of MR-based 4-PAM modulators, such as 4-PAM-SS, 4-PAM-EDAC, and 4-PAM-ODAC. We analyze how different types of MR-based 4-PAM modulators compare with the traditional OOK modulators at the photonic link-level and PNoC architecture-level while considering hardware overhead, performance, energy-efficiency, and reliability, and especially in the presence of inter-channel crosstalk. Our analysis shows that designing the constituent photonic links of PNoCs is subject to inherent tradeoffs among the achievable performance (aggregated datarate), energy consumption, and reliability, irrespective of the utilized modulation method and modulator type. Optimizing these design tradeoffs often involves finding the right balance between the photonic link's aggregated datarate and energy-reliability behavior. We find that different modulation methods and modulator types are differently positioned to achieve this balance: i.e., which modulation method and modulator type achieves better balance really depends on the underlying PNoC architecture.

Our novel contributions in this paper are as follows:
- We present an overview of how different MR-based 4-PAM modulators generate 4-PAM optical signals, and then compare their operation with a conventional MR-based OOK modulator;
- We present how the hardware implementation overheads for different 4-PAM modulation methods compare with one another, and with the conventional OOK modulation method ([78],[79]);
- We provide a systematic analysis of various design factors that affect the photonic link-level design tradeoffs for both OOK- and 4-PAM-based links;
- We utilize a heuristic-based search method to optimize the designs of DWDM-based photonic links with OOK, 4-PAM-SS, 4-PAM-EDAC, and 4-PAM-ODAC modulation methods, to achieve the desired balance between the aggregated datarate and energy-efficiency while achieving the BER of $10^{-9}$ or lower;
- We analyze how the optimized OOK and various 4-PAM photonic links affect the performance and energy-efficiency of two well-known PNoC architectures: CLOS PNoC [38] and SWIFT PNoC [8].

## 2 Background: Various Designs of OOK and 4-PAM Modulators From Prior Work

In this section, we present an overview of different MR-based OOK and 4-PAM modulator designs from prior work. In general, an MR-based modulator employs some mechanism to modulate the optical signal transmission at its through port (see Fig. 1(a)). In OOK modulators, the through-port optical transmission is modulated between two distinct levels, whereas for 4-PAM modulators it is modulated between four distinct levels. An MR-based modulator is fundamentally a wavelength-selective resonator whose employed modulation mechanism generally alters its resonant wavelength ($\lambda_r$) with respect to a utilized carrier (i.e., input) λ. This in turn alters the modulator's through-port optical transmission at the carrier λ. Most of the MR-based OOK and 4-PAM modulators (shown in Fig. 1 to Fig. 4) from prior work, e.g., [7], [21], [22], [36], [37], [39], utilize voltage biasing induced free-carrier injection/depletion, and the resultant free-carrier dispersion (FCD) mechanism [40], to modulate their through-port optical transmissions. However, different modular designs differ in their physical implementations, as a result, their area-energy-reliability footprints also differ. The following subsections present the operational details of different MR-based OOK and 4-PAM modulator designs and their physical implementations.

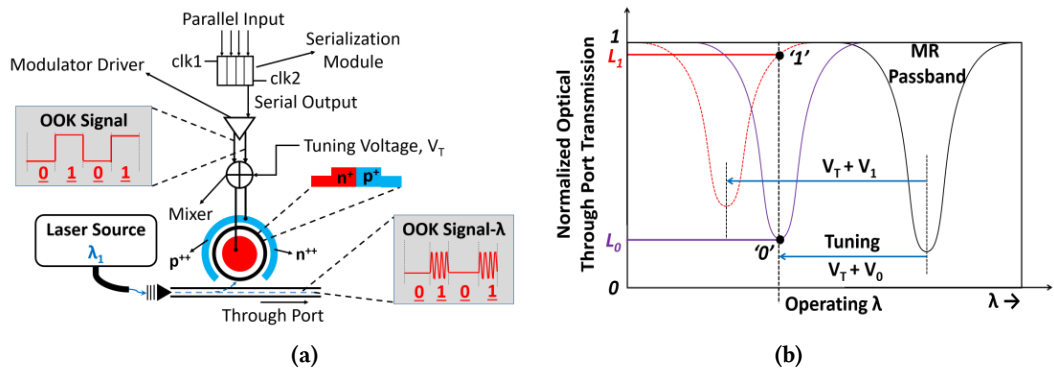

**Fig. 1: Illustration of (a) an MR-based on-off keying (OOK) modulator, and (b) the modulator's resonance passbands and optical transmission levels.**

## 2.1 MR-Based On-Off Keying (OOK) Modulator

Fig. 1(a) illustrates a typical MR-based OOK modulator [39], which employs a serialization module and a driver circuit that can produce a sequence of signal bias voltages corresponding to the input sequence of electrical bits (i.e., '1's and '0's). The modulator MR's resonance is switched in and out of alignment with signal- $\lambda_1$ by applying the sequence of signal-bias voltages to the MR. Before the MR modulator is driven by the signal bias voltages, each signal bias voltage in the sequence might be offset with a corresponding non-zero tuning-bias voltage to compensate for the resonant shift in the MR [18] that can occur because of the variations in the width and thickness of the MR during a conventional non-ideal fabrication process [54]. Such fabrication process related variations are referred to as process variations (PV) henceforth. Such resonant shift in the MR can also occur due to thermal variations (TV). For example, Fig. 1(b) illustrates how shifts in the resonance passband of an example OOK MR modulator can modulate it's through-port optical transmission. In Fig. 1(b), $V_T$ is the tuning bias voltage that depends on the magnitude of the PV-induced MR resonance misalignment, whereas $V_1$ and $V_0$ are input signal-bias voltages corresponding to logic '1' and logic '0' bits, respectively. Thus, from the figure, for the OOK MR modulator, the net-bias voltages of $V_T + V_1$ and $V_T + V_0$ yield, respectively, 'on' ($L_1$) and 'off' ($L_0$) levels of through-port optical transmission. As a result, an OOK MR modulator takes a sequence of bias voltages corresponding to data bits as input and generates an on-off keying (OOK) modulated optical signal as output.

## 2.2 MR-Enabled Signal Superposition Based 4-PAM Modulator (4-PAM-SS Modulator)

Fig. 2(a) illustrates a signal superposition based 4-PAM modulator design (referred to as 4-PAM-SS) for use in PNoCs, which was first proposed in [7]. From the figure, in a 4-PAM-SS modulator, two OOK MR modulators that are connected in parallel to two different waveguides generate two OOK-modulated optical signals of same λ but of different intensities in the ratio 2:1. These two OOK-modulated optical signals are superposed using a combiner to generate a 4-PAM modulated λ signal. As evident from Fig. 2(a), the need of an asymmetric power splitter and combiner can complicate the implementation of this design. This issue can be mitigated by using a different 4-PAM-SS design from [36] as shown in Fig. 2(b), which employs two cascaded OOK MR modulators coupled to a single waveguide to eliminate the need for a power splitter and combiner. Both these 4-PAM-SS modulator designs (Figs.

2(a) and 2(b)) in general require the two OOK-modulated optical signals to be in phase, which may not be possible to achieve under PV and TV (see Section 3.2.3) , potentially causing some reliability issues that will be discussed in Section 3. We utilize the 4-PAM-SS modulator design from [36] (Fig. 2(b)) for our analysis presented henceforth.

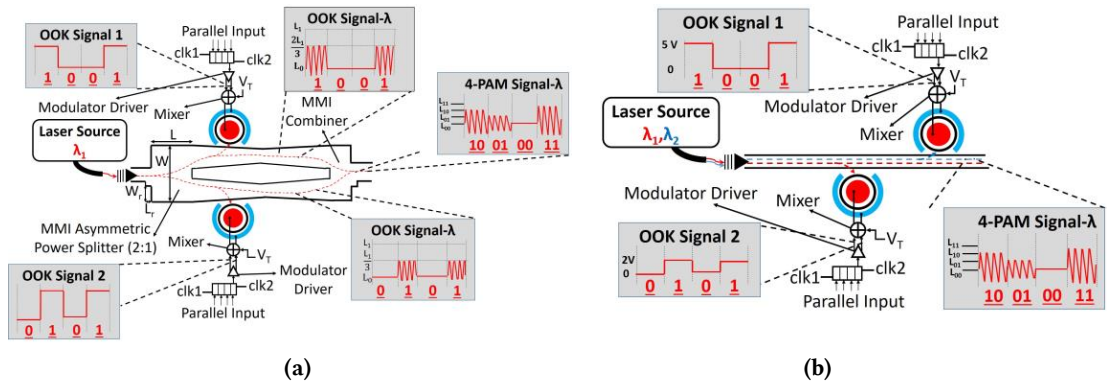

**Fig. 2: 4-PAM-SS modulator designs. (a) Design from [7] with two parallel OOK MR modulators and multi-mode interference (MMI) based asymmetric power splitter-combiner from [41]. (b) Design from [36] with two cascaded OOK MR modulators.**

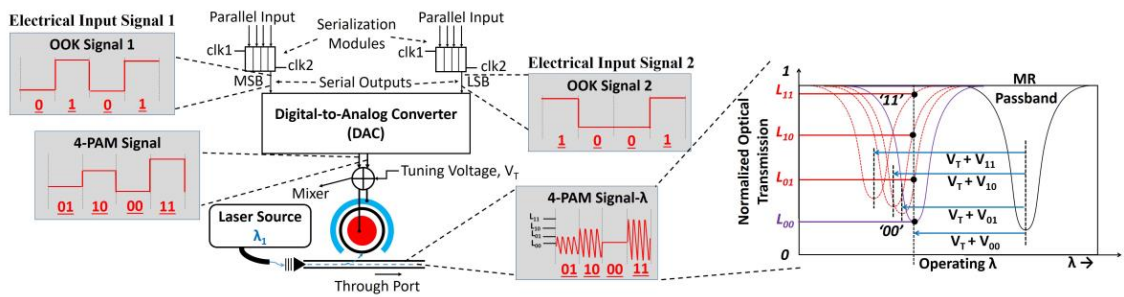

**Fig. 3: Illustration of an electrical DAC (EDAC) enabled MR-based 4-PAM modulator from [21]. Inset: Illustration of resonance passbands and optical transmission levels for an EDAC-enabled MR-based 4-PAM modulator.**

## 2.3 Electrical DAC (EDAC) Enabled MR-Based 4-PAM Modulator (4-PAM-EDAC Modulator)

In [21], an MR-based 4-PAM modulator is presented that utilizes an electrical DAC (EDAC) to convert two electrical OOK signals into an electrical 4-PAM signal, as shown in Fig. 3. This electrical 4-PAM signal is used by the driver circuit that drives an MR modulator to generate a proportional optical 4-PAM signal. The driver circuit generates four different bias voltages corresponding to the four distinct two-bit patterns (i.e., '00', '01', '10', '11') in the input electrical 4-PAM signal. These four voltages can induce four different optical transmission levels at the through port of the MR modulator, corresponding to four different magnitudes of resonance passband shift in the MR, as shown in Fig. 3 (see the inset). To achieve these transmission levels $L_{11}$, $L_{10}$, $L_{01}$, $L_{00}$ (shown in the inset of Fig. 3), the signal bias voltages $V_{00}$, $V_{01}$, $V_{10}$, $V_{11}$ of the modulator have to be decided upon appropriately. This can be done efficiently using the pulsed-cascode, a digital-to-analog converter (DAC), based output driver circuit reported in [21]. This circuit from [21] has a provision for sweeping the modulator bias voltages ($V_{10}$, $V_{01}$) to determine the target transmission levels ($L_{10}$, $L_{01}$) such that they are equidistant from $L_{11}$ and $L_{00}$, which allows for the in-situ corrections of any degree of aberrations in the transmission levels that can arise due to the fabrication process variation induced changes in the Q-factor and extinction ratio of the modulators. But this EDAC based 4-PAM signaling method incurs substantial area and energy overheads related to the required EDAC circuits [21], which can offset the general benefits of 4-PAM signaling.

## 2.4 Optical DAC (ODAC) Enabled MR-Based 4-PAM Modulator (4-PAM-ODAC Modulator)

To reduce the area and energy overheads of EDAC enabled MR modulators, an optical DAC (ODAC) enabled MR-based 4-PAM modulator was proposed in [22]. This modulator design consists of a spoked MR that functions like an ODAC to directly convert two input electrical OOK signals into a 4-PAM optical signal. A spoked MR is realized by segmenting its embedded P-N junction into multiple anodes and cathodes (e.g., 32 anodes and 32 cathodes in [22], and 15 anodes and 15 cathodes in the MR modulator shown in Fig. 4). All cathode segments are connected together via a spoked-ring shape metal contact in the center of the MR, while each anode segment has its own contact pin using which each anode segment can be driven independently or in some combination of other anode segments. For instance, in Fig. 4, a total of 10 out of 15 anode segments are connected and driven by electrical OOK signal 1, and the remaining 5 anode segments are driven by electrical OOK signal 2. This arrangement of the MR modulator's anode connections corresponds to four distinct spectral positions of the MR's resonance passband, which in turn corresponds to four distinct levels of optical transmission at the MR's through port. Thus, this spoked-MR-based modulator design functions like an ODAC to reduce the typical two-stage electro-optic OOK-to-4PAM conversion process to a single-stage process. Compared to the other MR-based 4-PAM modulators, this ODAC-enabled spoked-MR based 4-PAM modulator exhibits low area overhead and dynamic energy consumption [22].

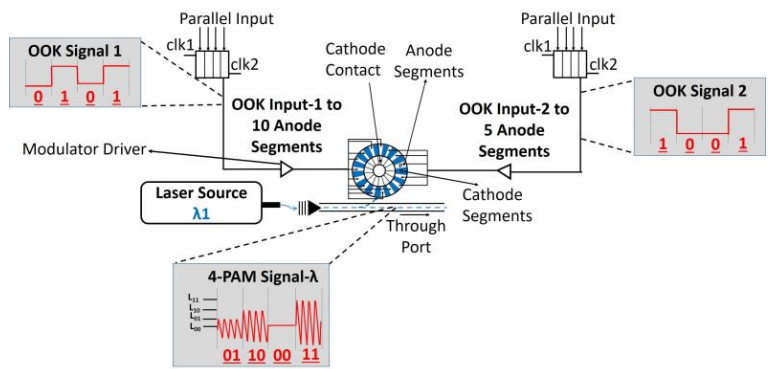

**Fig. 4: Optical DAC (ODAC) enabled MR-based 4-PAM signaling modulator from [22].**

*In summary*, different MR-based OOK and 4-PAM modulator designs function differently at the device level. Due to these functional differences, it can be intuitive inferred that different modulator designs would have different energy-performance behavior and implementation overheads at the link- and system-level. In the next section, we systematically analyze the physical-layer design overheads and static power consumption for various photonic link implementations that are based on different modulator designs and signaling methods discussed here.

## 3 Systematic Analysis of Photonic Links With Various Modulator Implementations

Recent advancements in CMOS-photonics integration (e.g., as demonstrated in ([76]-[79]) have enabled an exciting solution in the form of photonic network-on-chip (PNoC) architectures. Several PNoC architectures have been proposed till date ([8],[59],[64],[66]-[68]) that employ either fully optical interconnects or hybrid optical-electrical interconnects. In this section, we identify the physical-layer hardware components of PNoCs and their building blocks (i.e., photonic links), whose implementation overheads are highly affected by the choice of signaling method and modulator design.

Typically, a PNoC comprises of multiple photonic links. A photonic link comprises of one or more photonic waveguides, which move data packets between sender and receiver nodes in the optical domain over multiple DWDM wavelength channels. However, all data packet transfers outside of the PNoC in a manycore processor chip, e.g., between the processing cores and the PNoC, still occur in the electrical domain. Therefore, in a photonic link, it is important to enable electrical-to-optical (E/O) conversion of incoming data packets, which is typically achieved using a bank of MR-based modulators at the sender node. Similarly, to enable optical-to-electrical (O/E) conversion of outgoing data packets from a link, a bank of MR-based filters and photodetectors are employed at the receiver node. Both OOK and 4-PAM optical signaling based links require E/O conversion at the sender nodes and O/E conversion at the receiver nodes. The O/E converted signals at the output of photodetectors generally follow the format of the input optical signals, i.e., an OOK (4-PAM) modulated optical signal is converted into an OOK (4-PAM) modulated electrical signal by the photodetector. The same photodetector can be used to convert both OOK and 4PAM modulated optical signals to electrical signals. These photodetector output signals are generally reshaped by trans-impedance receiver modules to make them digitally processable.

Figs. 5(a) and 5(b) show the schematics of example trans-impedance receiver modules for OOK and 4-PAM signals, respectively. From the figures, the example 4-PAM receiver module employs three trans-impedance op-amps to generate two bit-streams, compared to the example OOK receiver that employs one trans-impedance op-amp to generate one bit-stream. The E/O and O/E conversion of signals in DWDM photonic links also utilize serialization and deserialization modules. At the E/O conversion unit of a DWDM photonic link, the converted optical data packets are transferred over different channels (i.e., each wavelength is an optical channel) at a higher bitrate than the bitrate of the incoming electrical data packets. To enable this conversion between bitrates, a serialization module is utilized before each MR-based modulator at the source node, and a deserialization module is used after each MR-based detector at the receiver node. Serialization modules can be implemented using parallel-in serial-out electronic buffers, whereas deserialization modules can be implemented using serial-in parallel-out electronic buffers, as shown in Figs. 5(c) and 5(d), respectively. From Fig. 5 and compared to OOK signaling, for a link with $N_\lambda$ wavelengths, using 4-PAM signaling (i.e., B = 2 in Fig. 5) requires 2× narrower electronic buffers in each (de)serialization module of the link. This is because using 4-PAM signaling in the link requires 2× number of (de)serialization modules compared to OOK signaling. As a result, for 4-PAM signaling, each incoming/outgoing data packet is striped across 2× number of electronic buffers (corresponding to 2× number of (de)serialization modules), allowing each buffer to be 2× narrower.

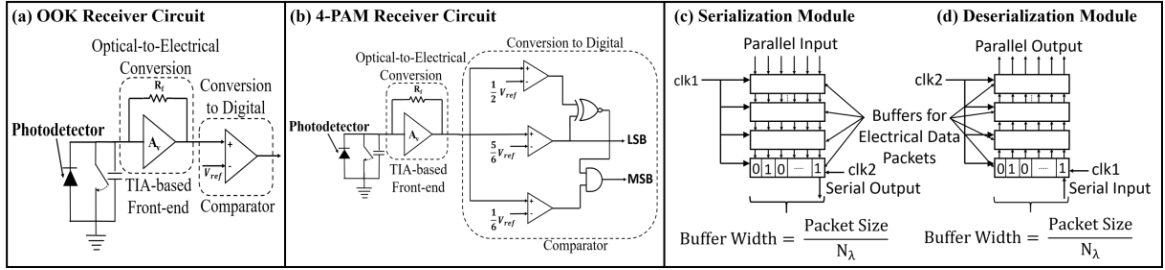

Fig. 5: Schematics of (a) a receiver module for an OOK modulation-based link [7], [22], (b) a receiver module for a 4-PAM modulation-based link [7][22], (c) a serialization module [42], and (d) a deserialization module [42]. $N_\lambda$ is the number of DWDM signals in the link. B is number of bits per symbol; B=1 for OOK signaling, and B=2 for 4-PAM signaling.

In summary, for a DWDM link, the overhead (e.g., area, power consumption) of implementing (de)serialization, and E/O and O/E conversion ultimately depends on the choice of signaling method and modulator design. This is because such a choice directly controls the required number of (de)serialization modules, number of MR-based modulators and filters, and the required type and count of photodetectors and receiver modules. Table 1 gives the number of required modules/instances of several hardware components (e.g., MR modulators, MR filters, photodetectors) for implementing DWDM photonic links with various signaling methods, such as OOK, 4-PAM-SS, 4-PAM-EDAC, and 4-PAM-ODAC. Table 2 gives example values (extracted from prior work) for dynamic energy consumption of several hardware components. The energy consumption value for a TIA based receiver front-end ($E^{TI\text{-}OPAMP}$) can change with the change in the technology node. But we do not expect $E^{TI\text{-}OPAMP}$ to affect the results provided in Tables 5 and 6, because $E^{TI\text{-}OPAMP}$ does not affect any of the link configuration parameters such as $N_\lambda$, $PP_{dB}$, S, or BR. Nevertheless, we point the reader to [86] for more detailed analysis of how the dynamic energy consumption of a TIA circuit changes for different design parameters of the circuit. Nevertheless, note that the study presented in this paper is independent of the parameter values provided in Table 1 and can be applied considering other parameter values. Moreover, we adopt the common research approach from prior works ([62]-[69]) and select the energy consumption values for various devices from different references (Tables 2 and 4) to undertake the link and system-level evaluations presented in this paper. We discuss the information provided in Table 1 in the following subsections (Subsection 3.1 to 3.4).

**Table 1. Number of instances, dynamic energy-per-bit (EPB), and static power values for various hardware components required for implementing a DWDM photonic link with $N_\lambda$ channels using various signaling methods and modulator types.** *PS* **is packet size in bits.**

| | Signaling Method/Modulator Type | | | |
|---|---|---|---|---|
| | | | 4-PAM | |
| Parameter | OOK | SS | EDAC | ODAC |
| #instances of various hardware components | | | | |
| # MR modulators | $N_\lambda$ | $2 \times N_\lambda$ | $N_\lambda$ | $N_\lambda$ |
| # MR filters | $N_\lambda$ | $N_\lambda$ | | |
| # Photodetectors | $N_\lambda$ | $N_\lambda$ | | |
| # Receiver modules | $N_\lambda$ | $N_\lambda$ | | |
| # Serialization modules | $N_\lambda$ | $2 \times N_\lambda$ | | |
| # Deserialization modules | $N_\lambda$ | $2 \times N_\lambda$ | | |
| # Buffer width in (de)-serialization modules (Fig. 5) | $PS/N_\lambda$ | $PS/(2 \times N_\lambda)$ | | |
| # Modulator drivers | $N_\lambda$ | $2 \times N_\lambda$ | $N_\lambda$ | $2 \times N_\lambda$ |
| # Total trans-impedance op-amps | $N_\lambda$ | $N_\lambda$ | | |
| # Total comparator op-amps | $N_\lambda$ | $3 \times N_\lambda$ | | |
| energy-per-bit (EPB) and static power values (45nm SOI-CMOS) | | | | |
| Total modulator driver EPB (pJ/bit) | $E^{Mod,OOK} \times N_\lambda$ | $E^{Mod,OOK} \times 2N_\lambda$ | $E^{Mod,EDAC} \times N_\lambda$ | $E^{Mod,ODAC} \times 2N_\lambda$ |
| Total serialization + deserialization EPB (pJ/bit) | $E^{SerDes} \times N_\lambda$ | $E^{SerDes} \times 2 \times N_\lambda$ | | |
| Total comparator op-amps EPB (pJ/bit) | $E^{CO\text{-}OPAMP} \times N_\lambda$ | $E^{CO\text{-}OPAMP} \times 3N_\lambda$ | | |
| Total trans-impedance op-amps EPB | $E^{TI\text{-}OPAMP} \times N_\lambda$ | $E^{TI\text{-}OPAMP} \times N_\lambda$ | | |
| Power of MR tuning control circuit (μW) | $P^{TC} \times 2N_\lambda$ | $P^{TC} \times 3N_\lambda$ | $P^{TC} \times 2N_\lambda$ | $P^{TC} \times 2N_\lambda$ |
| Microheater power (μW/nm) | $P^{\mu heater} \times 2N_\lambda$ | $P^{\mu heater} \times 3N_\lambda$ | $P^{\mu heater} \times 2N_\lambda$ | $P^{\mu heater} \times 2N_\lambda$ |

## 3.1 Photonic Links Based on OOK Modulation

Fig. 6(a) shows a schematic of an OOK signaling based DWDM link with four optical channels ($N_\lambda$ = 4). From the figure, the link utilizes four instances of MR modulators, modulator drivers, MR filters, photodetectors, receiver modules, serialization modules, and deserialization modules each. Therefore, one can generalize that for a DWDM OOK link with $N_\lambda$ channels, it would require $N_\lambda$ instances of each of the various hardware components as mentioned in Table 1. Moreover, the link uses one trans-impedance op-amp per receiver module (Fig. 5(a)), requiring $N_\lambda$ trans-impedance op-amps corresponding to $N_\lambda$ receiver modules (Table 1). Lastly, having a total of $N_\lambda$ (de)serialization modules per link leads to each buffer being of size (Packet Size/$N_\lambda$) bits wide (Figs. 5(c) and 5(d)), as B=1 for OOK links in Figs. 5(c) and 5(d). Moreover, the link has total energy-per-bit (EPB) and static power consumption values associated with various hardware components (see Table 2).

Table. 2 Sample values of per-instance EPB for modulator driver ($E^{Mod}$), serialization + deserialization ($E^{SerDes}$), trans-impedance op-amps ($E^{TI\text{-}OPAMP}$) and per-MR static power for MR tuning control circuit ($P^{TC}$) and microheater ($P^{\mu heater}$).

| Parameter | Signaling Method | | |
|---|---|---|---|
| | OOK (pJ/bit) | 4-PAM EDAC (pJ/bit) | 4-PAM ODAC (pJ/bit) |
| $E^{Mod}$ | 0.13 [43] | 3.04 [21] | 0.04 [22] |
| $E^{SerDes}$ | 0.5 pJ/bit [22] | | |
| $E^{CO\text{-}OPAMP}$ | 0.21 pJ/bit [43] | | |
| $E^{TI\text{-}OPAMP}$ | 0.24 pJ/bit [86] | | |
| $P^{TC}$ | 385 µW [44] | | |
| $P^{\mu heater}$ | 800 µW/nm [45] | | |

From Table 2, a typical OOK modulator driver consumes EPB of $E^{Mod,OOK}$ = 0.13 pJ/bit [43], a typical serialization and deserialization module consumes EPB of $E^{SerDes}$ = 0.5 pJ/bit [22], and a typical trans-impedance op-amp consumes EPB of $E^{TI\text{-}OPAMP}$ = 0.21 pJ/bit [43]. As a result, an OOK link with $N_\lambda$ channels consumes modulator driver EPB of ($E^{Mod,OOK} \times N_\lambda$) pJ/bit, serialization + deserialization EPB of ($E^{SerDes} \times N_\lambda$) pJ/bit, and trans-impedance op-amps EPB of ($E^{TI\text{-}OPAMP} \times N_\lambda$) pJ/bit, as the link has $N_\lambda$ counts of modulator drivers, serialization modules, deserialization modules, and trans-impedance op-amps each. Further, from Table 2, the tuning control circuit and the integrated microheater of an MR consume $P^{TC}$ = 385µW [44] and $P^{\mu heater}$ = 800µW/nm [45] power, respectively. Therefore, the OOK link consumes ($P^{TC} \times 2 \times N_\lambda$) µW power for the MR tuning control circuits and ($P^{\mu heater} \times 2 \times N_\lambda$) µW/nm power in the MR-integrated microheaters, as the link has $2 \times N_\lambda$ MRs ($N_\lambda$ modulators + $N_\lambda$ filters).

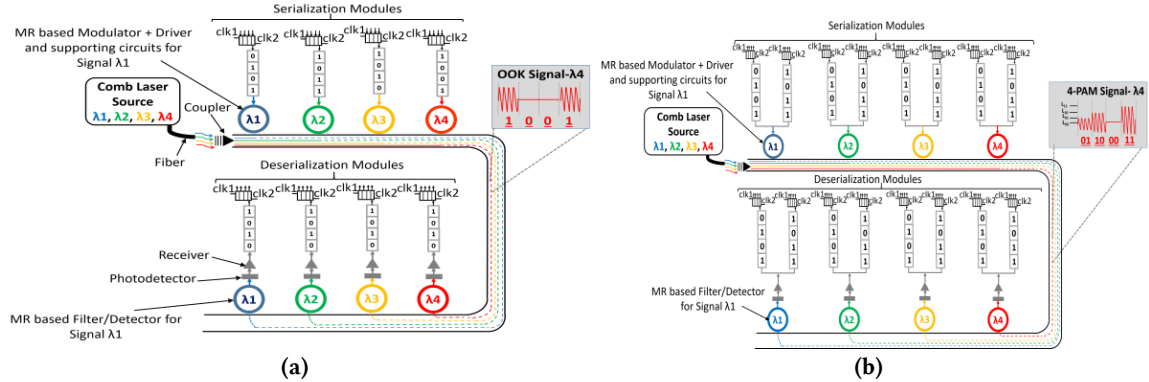

**Fig. 6:** Schematic illustration of (a) an OOK modulation-based optical link, and (b) a 4-PAM modulation-based optical link, with total four wavelengths ($\lambda_1$ to $\lambda_4$). Note that having equal number of optical signals results in 2× datarate for the 4-PAM link. In other words, equal datarate can be achieved for 4-PAM links by using 2× less optical signals.

### 3.2 DWDM Links Using 4-PAM Signaling and Various 4-PAM Modulators

Fig. 6(b) shows a schematic of a 4-PAM signaling based DWDM link with four optical channels ($N_\lambda$ = 4). It is evident from the figure that, compared to an OOK link, a 4-PAM link with $N_\lambda$ = 4 requires 2× more serialization and deserialization modules. Moreover, a 4-PAM receiver requires 3× more trans-impedance op-amps (Fig. 5). Therefore, it can be generalized that contrary to an OOK link, a 4-PAM link with $N_\lambda$ channels requires 2×$N_\lambda$ serialization modules, 2×$N_\lambda$ deserialization modules, 3×$N_\lambda$ trans-impedance op-amps based receiver modules (Table 1). Moreover, a 4-PAM link requires (Packet Size/(2×$N_\lambda$)) bits of buffer width at their E/O and O/E interfaces (Table 1), as the number of bits per symbol B=2 for a 4-PAM link in Figs. 5(c) and 5(d). On the other hand, like an OOK link, the 4-PAM link also requires $N_\lambda$ counts of MR filters, photodetectors, and receiver modules each.

Corresponding to these hardware component counts, a 4-PAM link with $N_\lambda$ channels consumes a total serialization and deserialization EPB of ($E^{SerDes} \times 2 \times N_\lambda$) pJ/bit and total trans-impedance op-amps EPB of ($E^{TI-OPAMP} \times 3 \times N_\lambda$) pJ/bit (Table 1). In addition, the counts and overheads of other hardware components for a 4-PAM link, such as MR modulators and modulator drivers, depend on the specific 4-PAM modulator type, as discussed next.

### 3.2.1 4-PAM EDAC Modulator Based Links

A 4-PAM-EDAC modulator-based link with $N_\lambda$ channels requires a total of $N_\lambda$ MR modulators, which makes the total MRs per link to be $2 \times N_\lambda$ ($N_\lambda$ filters and $N_\lambda$ modulators). Moreover, the link requires $N_\lambda$ electrical DAC (EDAC) based modulator drivers (one EDAC per modulator as shown in Fig. 3), each of which consumes $E^{Mod,EDAC} = 3.04$ pJ/bit EPB [21]. Therefore, a 4-PAM-EDAC link with $N_\lambda$ channels consumes modulator driver EPB of ($E^{Mod,EDAC} = 3.04 \times N_\lambda$) pJ/bit, power for MR tuning control circuits of ($P^{TC} \times 2 \times N_\lambda$) μW, and MR microheater power of ($P^{\mu heater} \times 2 \times N_\lambda$) μW/nm (Table 1).

### 3.2.2 4-PAM ODAC Modulator Based Links

A 4-PAM-ODAC modulator-based link with $N_\lambda$ channels requires a total of $N_\lambda$ spoked MR modulators (Fig. 4), which makes the total number of MRs per link to be $2 \times N_\lambda$ ($N_\lambda$ filters and $N_\lambda$ modulators). However, unlike a 4-PAM-EDAC link, a 4-PAM-ODAC link requires $2 \times N_\lambda$ modulator drivers (2 drivers per modulator; Fig. 4), each of which consumes $E^{Mod,ODAC} = 0.04$ pJ/bit EPB [22]. Therefore, a 4-PAM-ODAC link with $N_\lambda$ channels consumes modulator driver EPB of ($E^{Mod,ODAC} = 0.04 \times 2 \times N_\lambda$) pJ/bit, power for MR tuning control circuits of ($P^{TC} \times 2 \times N_\lambda$) μW, and MR microheater power of ($P^{\mu heater} \times 2 \times N_\lambda$) μW/nm (Table 1).

### 3.2.3 4-PAM-SS Modulator Based Links

A 4-PAM-SS modulator-based link with $N_\lambda$ channels requires $2 \times N_\lambda$ MR modulators (2 modulators per channel; Fig. 2), which makes the total number of MRs per link to be $3 \times N_\lambda$ ($N_\lambda$ filters + $2 \times N_\lambda$ modulators). Moreover, a 4-PAM-SS link requires $2 \times N_\lambda$ modulator drivers (1 driver per modulator; Fig. 2), each of which utilizes OOK signaling and consumes $E^{Mod,OOK} = 0.13$ pJ/bit EPB [43]. Therefore, a 4-PAM-SS link with $N_\lambda$ channels consumes in total modulator driver EPB of ($E^{Mod,OOK} \times 2 \times N_\lambda$) pJ/bit, power for MR tuning control circuits of ($P^{TC} \times 3 \times N_\lambda$) μW, and MR microheater power of ($P^{\mu heater} \times 3 \times N_\lambda$) μW/nm (Table 1).

In addition, a 4-PAM-SS link also suffers from a high signal power loss in 4-PAM modulators due to the possible inter-channel crosstalk. Ideally, in a 4-PAM-SS modulator, when two OOK-modulated signals are superposed (Fig. 2), a 4-PAM modulated signal is generated owing to the constructive interference between the two OOK signals. However, the constructive interference happens only if both OOK signals are in phase. Unfortunately, in the presence of non-idealities such as fabrication process and on-chip temperature variations, a significant phase difference may exist between the two superposed OOK signals, which can lead to destructive interference between them. Owing to the random nature of fabrication-process variations, this incurred phase difference may fall anywhere in the range from 0 to $2\pi$. This implies that the degree of destructive interference incurred between the OOK signals due to the phase difference (and hence the amplitude levels of the symbols of the resultant 4-PAM signal) may fall anywhere in a very large range of values. This in turn makes it very hard to ensure reliability of communication with a 4-PAM-SS modulator. We evaluate the adverse impact of random fabrication-process variations on the reliability of 4-PAM-SS links in terms of the worst-case destructive interference, as explained next.

The worst-case destructive interference in a 4-PAM-SS modulator occurs when the two superposed OOK signals are completely out of phase, i.e., when the phase difference between them is an odd multiple of $\pi$. The amount of signal loss due to the superposition of two out-of-phase OOK signals depends on their individual signal intensities. Typically, as explained in Section 2.2 (Fig. 2), in a 4-PAM-SS modulator, to equidistantly space the four amplitude levels of the output 4-PAM symbol in the available range of optical transmission, the intensities of the individual OOK signals are kept at two-third and one-third of the intensity of the conventional OOK signal. Hence, for the best-case constructive interference between the superposed OOK signals, the intensity of the resultant 4-PAM signal becomes $2/3 + 1/3 = 1$. In contrast, for the worst-case destructive interference, the intensity of the resultant 4-PAM signal becomes $2/3 - 1/3 = 1/3$, causing the worst-case interference-related signal loss to be $-10 \times \log(1/3) = 4.8$dB. This interference-related signal loss in 4-PAM-SS modulators reduces the signal-to-noise ratio (SNR) and increases the bit-error rate (BER). We have considered this worst-case interference-related signal loss of 4.8dB in our tradeoff analysis for 4-PAM-SS links. We have also considered the best-case scenario for which this interference-related loss is omitted for our analysis of 4-PAM-SS links (see Sections 4.1.2, 4.1.4, and 4.2). This in turn reduces the overall communication reliability, adversely affecting the tradeoffs among the energy, reliability, and performance of 4-PAM-SS links.

*In summary*, the hardware overhead of implementing an OOK or 4-PAM signaling based photonic link depends not only on the choice of modulator design and signaling method but also on the number of parallel wavelength channels $N_\lambda$ in the link (see Table 1). The maximum supportable $N_\lambda$ for an OOK or 4-PAM signaling based photonic link is determined based on the inherent tradeoffs among the energy consumption, reliability (BER), and performance of the designed OOK or 4-PAM link, as discussed in the next section.

## 4 Design Tradeoffs for Photonic Links

Designing a photonic link is subject to inherent tradeoffs among the achievable performance (aggregated datarate), energy consumption, and reliability (BER). Optimizing these design tradeoffs often involves finding the balance between the link's aggregated datarate and energy-reliability behavior. From [6], [10], [13], and [33], for optimizing the design of a photonic link, optical power budget ($P^B_{dB}$) of the link is the most critical design constraint. It is calculated in dB as the difference between the maximum allowable optical power ($P_{Max}$) and detector sensitivity ($S$), as shown in Eq. (1). For a photonic link, $P_{Max}$ identifies the ceiling of $P^B_{dB}$ and ensures that the total power of all the DWDM signals (i.e., total $N_\lambda$ signals) propagating through the link remains below the maximum allowable level which is limited by various non-linear effects of silicon in constituent devices [33]. On the other hand, $S$ is the noise-limited floor of the link's $P^B_{dB}$ and ensures that the individual signals propagating through the link reach the receiver without dropping below the minimum power level defined by $S$. For a photonic link, the total optical power allocated within its $P^B_{dB}$ supports two causes, as shown in Eq. (2). First, it compensates for the total optical power penalty ($PP_{dB}$) of the link. Second, it supports total $N_\lambda$ DWDM wavelength signals/channels in the link. Therefore, from Eq. (2), improving the performance (aggregated datarate) of a photonic link that has a fixed $P^B_{dB}$ by increasing the supported $N_\lambda$ in the link requires a corresponding decrease in the link's $PP_{dB}$.

$$P^B{}_{dB} = P_{Max} - S \qquad (1)$$

$$P^B{}_{dB} \geq PP_{dB} + 10\log_{10}(N_\lambda) \qquad (2)$$

For optimal link design, this tradeoff between $N_\lambda$ and $PP_{dB}$ is affected by the following four factors: *(i)* The photodetector sensitivity $S$, which is the minimum detectable optical power in dBm, and depends on the baud-rate (i.e., number of amplitude/level transitions in unit time) of the individual photonic signals [33]. *(ii)* The cyclic dependency between $PP_{dB}$ and achievable aggregated data rate, which is given as $N_\lambda$ × bitrate ($BR$) of photonic channels. This cyclic dependency means that $PP_{dB}$ depends on $N_\lambda \times BR$ through the available $P^B_{dB}$, whereas $N_\lambda \times BR$ in turn depends on $PP_{dB}$. *(iii)* Several spectral parameters of MRs such as the free-spectral-range (FSR) and full-width-at-half-maximum (FWHM) bandwidths, which impact the effective value of $PP_{dB}$. *(iv)* The ultimate design goals of photonic links and PNoCs, including the goals of maximizing aggregated datarate, energy-efficiency, and/or achieving the desired BER, which also impact the effective value of $PP_{dB}$. In the next subsection, we systematically analyze and provide detailed models for all four aforementioned factors that affect photonic link design tradeoffs, with respect to the utilized modulator types and signaling methods.

### 4.1 Factors that Affect Photonic Link Design Tradeoffs

#### 4.1.1 Baud-Rate Dependent Detector Sensitivity (S)

From [33], detector sensitivity (S) in dBm increases with increase in signal baud-rate. Signal baud-rate is defined as the number of amplitude/level transitions in the photonic signal occurring in unit time. We consider the baseline value of S = -22dBm at 10Gb/s [33] for both 4-PAM and OOK links, and adopt the model from [33] to capture how S would increase for baud-rates greater than 10Gb/s. We extract S for 4-PAM signals based on the experimentally demonstrated and validated models from [20]. From [20], it is evident that a 4-PAM signal requires 3.3dB more received power compared to an OOK signal of the same bitrate (BR), to achieve the same bit-error rate (BER) as achieved by the OOK signal. Therefore, to derive S for 4-PAM signals, we simply add 3.3 dB to the S that is obtained for OOK signals of the same BR using the BR-dependent model of S from [33]. The same value of baud-rate translates into 2× bitrate for a 4-PAM link compared to an OOK link, and for evaluating link performance, bitrate (i.e., aggregated datarate) is a more useful metric than baud-rate. Therefore, we use the following Eq. (3) as the relation between baud-rate and bitrate, henceforth.

$$BaR = BR/(M/2) \tag{3}$$

Here, $BaR$ is baud-rate, $BR$ is bitrate, and $M$ is number of amplitude levels used in the signal to represent a symbol ($M$=2 for an OOK signal, and $M$=4 for a 4-PAM signal).

### 4.1.2 Cyclic Dependency Between PP$_{dB}$ and Aggregated Datarate (N$_\lambda$×BR)

To understand the cyclic dependency between $PP_{dB}$ and aggregated datarate (N$_\lambda$×$BR$), it is important to understand what constitutes $PP_{dB}$ (i.e., the total optical power penalty of the link) and how it changes between OOK and 4-PAM links. For a link, $PP_{dB}$ is comprised of the total penalty of the MR filter array ($PP^{Fil}$), MR modulator array crosstalk penalty ($PP^{Mod}$), PAM signaling penalty ($PP^{PAM}$), and various optical signal power losses such as waveguide propagation loss ($P_L^{WGP}$), waveguide bending loss ($P_L^{WGB}$), through loss of active MRs ($P_L^{MR-Act}$), through loss of inactive MRs ($P_L^{MR-InAct}$), worst-case signal interference penalty ($PP^{INTRF}$), and splitter/coupler loss ($P_L^{SpC}$). In [34], [27], and [24], $PP^{Fil}$ and $PP^{Mod}$ are analytically modeled, considering the general case of a bank of N$_\lambda$ modulator MRs employed at the sender node and a bank of N$_\lambda$ filter MRs employed at the receiver node of a link with N$_\lambda$ DWDM signals. Accordingly, $PP^{Mod}$ for an MR modulator in the bank of N$_\lambda$ modulator MRs can be evaluated using Eq. (4) and (5) [34]. Similarly, $PP^{Fil}$ for the ith MR in the bank of N$_\lambda$ MR filters can be given as Eq. (6) [27], where formulas for some important terms in Eq. (6) are given in Eqs. (7)-(9) [27]. Here, Eq. (7) considers the total crosstalk contribution from all N$_\lambda$ wavelength signals combined [27]. The definitions and typical values (if any) of various terms used in Eqs. (1)-(9) are given in Table 3 and Table 4, respectively.

$$PP^{Mod} = -5 * \log_{10}\left(\frac{\left(\frac{2K}{FWHM}\right)^2 + q_0}{\left(\frac{2K}{FWHM}\right)^2 + 1}\right) \tag{4}$$

$$K = \begin{cases} f_\Delta - \Delta f, & f_\Delta > 0 \\ f_\Delta, & f_\Delta < 0 \end{cases} \tag{5}$$

$$(PP^{Fil})_{i_{th}MR} = \left(-10 * \log_{10}\left(1 - 0.5 * Q_{BER} * \frac{P_{Xtalk}}{P_{NRZ}^{av}} * \frac{r+1}{r-1}\right)\right) \tag{6}$$

$$\left(\frac{P_{Xtalk}}{P_{NRZ}^{av}}\right)_{i_{th}MR} = \sum_{j=1, j \neq i}^{N_\lambda} \Gamma_{i,j} \tag{7}$$

$$\Gamma_{i,j} = \left(\int_{-\infty}^{+\infty} \frac{\text{sinc}^2(F) \, dF}{1 + \left(\frac{F+(j-i)F_\Delta}{\xi_i}\right)^2} * \left[\prod_{k=1}^{i-1} \frac{\left(\frac{F+(j-k)F_\Delta}{\xi_k}\right)^2}{1 + \left(\frac{F+(j-k)F_\Delta}{\xi_k}\right)^2}\right]\right) \tag{8}$$

$$(j-i)F_\Delta = \left(\frac{v_{Si}}{\lambda_i \times BaR} - \frac{v_{Si}}{\lambda_i + \left((j-i) \times \frac{FSR \, (nm)}{N_\lambda + 1}\right) \times BaR}\right) \tag{9}$$

Table 3: Definitions of various link design parameters and notations from Eqs. (1)-(15).

| Parameter | Definition |
|---|---|
| $P_{Max}$ | Max. allowable optical power in waveguide (dBm) [30] [33] |
| S | Detector sensitivity at 10Gb/s [33] (dBm) |
| $P_L^{WGB}$ | Waveguide bending loss (dB per 90°) [6] |
| $v_{Si}$ | Speed of light at 1550 nm in silicon (in m/s) |
| $Q_{BER}$ | Signal Q-parameter for BER = $10^{-9}$ [27] |
| $q_0$ | Extinction ratio of MR modulator in the OFF state [34] |
| $P_L^C$ | Coupler loss (dB) [51] |
| $P_L^{Sp}$ | Splitter Loss (dB) |
| $P_L^{WBP}$ | Propagation loss (dB), @ 1dB/cm [46] |
| $PP^{INTRF}$ | Worst-case signal interference penalty (dB) |
| $PP^{PAM}$ | 4-PAM signaling penalty (dB) |
| r | Extinction ratio of modulation |
| FWHM | 3dB Bandwidth of an MR modulator (GHz) |
| $PP^{ER}$ | Penalty (dB) due to the finite r (see above) [34] |
| $P^B_{dB}$ | Photonic link power budget in dB |
| $PP_{dB}$ | Total power penalty for the photonic link in dB |
| BR | Bitrate of a photonic signal |
| BaR | Baud-rate (# of level transitions per unit time) of a signal |
| M | # of amplitude levels per symbol in a photonic signal |
| $PP^{Mod}$ | Crosstalk power penalty in a modulator MR (in dB) |
| $f_\Delta$ | Frequency spacing between two adjacent photonic signals |
| $\Delta f$ | Frequency spacing between an MR modulator's OFF-state and ON-state resonances |
| $P_{NRZ}^{av}$ | Average power per incoming photonic signal at the i[th] MR filter in an MR filter bank |
| $P_{Xtalk}^{av}$ | Cumulative crosstalk power from all $N_\lambda$ signals combined at the i[th] MR filter of an MR filter bank |
| $PP^{Fil}$ | Crosstalk power penalty (in dB) at the i[th] MR filter |
| $\Gamma_{i,j}$ | Fraction of crosstalk power from j[th] signal dropped at the i[th] MR filter in an MR filter bank |
| F | Photonic signal frequency normalized to baud-rate (BaR) |
| $(j-i)F_\Delta$ | Frequency spacing between the i[th] MR filter resonance and j[th] signal normalized to baud-rate (BaR) |
| $\xi_i$ | FWHM of i[th] MR filter normalized to BaR |
| $\lambda_i$ | Resonance wavelength of i[th] MR filter |
| FSR | Free-spectral range |
| $N_\lambda$ | Number of photonic signals DWDM in a waveguide |
| $P_L^{MR-Act}$ | Through loss of an active MR |
| $P_L^{MR-InAct}$ | Through loss of an inactive MR |
| $P_{dB}^{BERO}$ | Power penalty (dB) for reliability optimal design of a link |
| $P_{dB}^{DR-BER-Bal}$ | Power penalty (dB) for datarate-reliability balanced link design |

Although Eqs. (4) to (9) were originally developed in [27] and [34] for OOK links, this same set of equations can be used to determine $PP^{Mod}$ and $PP^{Fil}$ for 4-PAM links as well. This is because OOK signals and 4-PAM signals have similar frequency spectra (i.e., in the shape of the sinc function). As a result, the utilized equations can be transformed to be based on signal baud-rate (BaR) instead of bitrate (BR), because the crosstalk at the modulators and filters can be assumed to have a Gaussian distribution, as demonstrated in [27]. Further, the parameters FWHM, $Q_{BER}$, and r in Eqs. (4) and (6) assume different values for OOK and 4-PAM signaling types (as shown in Table 4). Moreover, as a 4-PAM signal has 3× less separation between its amplitude levels compared to an OOK signal, a 4-PAM signal requires ~3.3dB more power at the receiver [22], compared to an OOK signal of the same BaR, to achieve the same bit-error rate (BER) of $10^{-9}$. This extra required power is accounted for as $PP^{PAM}$ (Table 4) in the total $PP_{dB}$ value. Moreover, note that to evaluate $PP^{Mod}$ for 4-PAM-SS links, we treat the 2 MR modulators required per wavelength signal as a single modulator unit (constituting a bank of $N_\lambda$ MR modulator units at the sender node), and $PP^{Mod}$ is evaluated for each MR modulator unit instead of each individual MR modulator. Also, using 8-PAM/16-PAM signals in the links can certainly reduce the hardware requirement for the links compared to 4-PAM signals, if the target aggregate datarate remains unchanged. This in turn can result in higher dynamic energy efficiency for 8-PAM/16-PAM links. However, PP$^{PAM}$ for 8-PAM and 16-PAM photonic links would increase to 6.1 dB and 8.75 dB respectively [20], compared to PP$^{PAM}$ of 3.3 dB for 4-PAM links (Table 4), due to the larger values of M for 8-PAM/16-PAM links (M = 8, 16 for 8-PAM, 16-PAM links respectively [20]). Larger PP$^{PAM}$ would increase the overall penalty PP$^{dB}$ for 8-PAM/16-PAM links, which in turn can render lower N$_\lambda$, lower BR, and hence, lower aggregate datarate to 8-PAM/16-PAM links, compared to 4-PAM links. This reduced aggregate datarate can offset the benefits obtained from achieving higher dynamic energy efficiency.

$$PP_{dB} = P_L^{MR-Act} + P_L^{MR-InAct} + P_L^{WGP} + P_L^{Sp} + P_L^{INTRF} + PP^{Mod} + PP^{Fil} + PP^{PAM} + P_L^{WGB} + P_L^C + PP^{ER} \quad (10)$$

$$P_{L, j_{th}\lambda}^{MR} = -10 \log_{10}\left(\sum_{i=1, j \neq i}^{N_\lambda} \Gamma_{i,j}\right) \quad (11)$$

$$(j-i)F_\Delta^{MR-Act} = \left(\frac{v_{Si}}{\lambda_i \times BaR} - \frac{v_{Si}}{\lambda_i + \left((j-i) \times \frac{FSR\ (nm)}{N_\lambda + 1}\right) \times BaR}\right) \quad (12)$$

$$(j-i)F_\Delta^{MR-InAct} = \left(\frac{v_{Si}}{\lambda_i \times BaR} - \frac{v_{Si}}{\lambda_i + \left((j-i+0.5) \times \frac{FSR\ (nm)}{N_\lambda + 1}\right) \times BaR}\right) \quad (13)$$

The values of *PP$^{Mod}$* and *PP$^{Fil}$* (as evaluated from Eqs. (4) to (9)), along with *PP$^{PAM}$*, contribute to *PP$_{dB}$*, as shown in Eq. (10). Eq. (10) also has some other terms related to the optical signal power loss. The definitions and typical values (if any) of all these loss terms from Eq. (10), except *P$_L^{MR-Act}$* and *P$_L^{MR-InAct}$* (which are discussed in the next paragraph), are also given in Table 3 and Table 4 respectively. Note that the values from Table 4 for some of the terms in Eq. (10) depend on the underlying signaling/modulator type and/or PNoC architecture. For example, from Eq. (10) and Table 4, *PP$^{INTRF}$* is zero for 4-PAM-ODAC, 4-PAM-EDAC, and OOK links, whereas it is 4.8dB for 4-PAM-SS links. This is because only 4-PAM-SS modulators incur signal superposition induced interference loss, as explained in Section 3.2. Similarly, the values of *r*, *FWHM*, and *PP$^{ER}$* also change between different modulator/signaling types (Table 4). Moreover, the values of *P$_L^{Sp}$* and *P$_L^{WGP}$* depend on the underlying PNoC architecture—we use CLOS [38] and SWIFT PNoC [8] PNoC architectures in this paper—as the required count of splitters and waveguide lengths differ between CLOS and SWIFT PNoCs.

It is evident from Eqs. (1)-(13) that there is a *cyclic dependency* between *PP$_{dB}$* and N$_\lambda$ as the achievable *N$_\lambda$* for a link depends on *PP$_{dB}$* from Eq. (1), whereas *PP$_{dB}$* in turn is determined based on the combination of N$_\lambda$ and bit-rate (*BR)* from *BaR* in *F$_\Delta$* in Eq. (4), (5), (8), (9), (12), (13). *This cyclic dependency makes it difficult to find out the optimal combination of N$_\lambda$ and BR that can be supported by a link.* To mitigate this problem, we employ a heuristic-based search approach that finds out the optimal combination of N$_\lambda$ and *BR*, as described in Section 4.2.

**Table 4: Typical values (if any) of various link design parameters and notations from Eqs. (1)-(15).**

| Parameter | Value | | | |
|---|---|---|---|---|
| $P_{Max}$ | 20 dBm [30][33] | | | |
| S | -22.5 dBm [33] @ 10 Gb/s BaR | | | |
| $P_L^{WGB}$ | 0.005 dB per 90° [6] | | | |
| $v_{Si}$ | $8.6 \times 10^7$ m/s | | | |
| $q_0$ | 0.04 [34] | | | |
| $P_L^C$ | 0.9 dB [51] | | | |
| | **PNoC Architectures** | | | |
| | CLOS | SWIFT | | |
| $P_L^{Sp}$ | 5.6 dB [38] | 1.2 dB [8] | | |
| $P_L^{WGP}$ at 1 dB/cm [46] | 4.5 dB (4.5 cm long link) [28] | 12 dB (12 cm long link) [8] | | |
| | **Signaling Methods** | | | |
| | OOK | SS | EDAC | ODAC |
| $PP^{INTRF}$ | 0 dB | 4.8 dB | 0 dB | 0 dB |
| $PP^{PAM}$ | 0 dB | | 3.3 dB [22] | |
| r | 5 dB [22] | 5 dB [22] | 5 dB [22] | 2 dB [22] |
| FWHM | 30 GHz [47] | 45 GHz [37] | 18 GHz [21] | 36 GHz [22] |
| $PP^{ER}$ | 4.2 dB [34] | 4.2 dB [34] | 4.2 dB [34] | 7.7 dB [34] |
| $Q_{BER}$ | 6 dB [27] | 12.5 dB [27], [12] | | |

### 4.1.3 Dependence of $PP_{dB}$ on MRs' Spectral Parameters

In Eq. (10), $PP_{dB}$ depends on the MRs' spectral parameters such as FWHM and FSR. Parameters FWHM and FSR are defined in Table 3. These spectral parameters depend on the device dimensions that are utilized for implementing the MRs of the photonic links. We select different FWHM values for MR modulators and filters based on the utilized modulator/signaling type, as shown in Table 4. Moreover, for FSR considerations, we select a viable FSR value of 20 nm from prior work [48] for our analysis in this paper (see Section 4.2). Our design methodology, analysis, and related link-level and system-level evaluation results are discussed in Section 4.2 and Section 5.

### 4.1.4 Dependence of $PP_{dB}$ on Design Goals

Whether or not to consider $PP^{Fil}$, $PP^{Mod}$, and $PP^{INTRF}$ in Eq. (10) to evaluate the effective value of $PP_{dB}$ depends on whether the goal is to design photonic links and PNoCs with maximum aggregated datarate or desired bit-error rate (BER). The emanation of crosstalk noise in modulator and filter MR banks reduces the signal-to-noise ratio (SNR) in photonic links (e.g., [6], [11]), which in turn increases the BER, degrading the reliability of photonic communication. To compensate for this degradation in BER, one way is to increase the input signal power by an appropriate amount. The required increase in the input signal power to achieve the unchanged BER in the presence of crosstalk noise is termed as power penalty. In Eq. (10), $PP^{Fil}$ and $PP^{Mod}$ correspond to the crosstalk noise induced power penalties for the filter MR bank and modulator MR bank, respectively. Similarly, $PP^{INTRF}$ corresponds to the required increase in the input signal power (i.e., caused power penalty) to compensate for the worst-case destructive signal interference in 4-PAM-SS links. From Table 4, our considered models and resultant values of $PP^{Fil}$ and $PP^{Mod}$ correspond to a BER of $10^{-9}$. We select BER of $10^{-9}$ for our analysis, because it is often considered acceptable for optical communication links [20], [34]. From this value of BER, we calculate $Q_{BER}$ (defined in Table 4) using the models presented in [27]. From Table 4, the evaluated $Q_{BER}$ differs between OOK and PAM4 signaling/modulation techniques. The presence of the $PP^{Fil}$, $PP^{Mod}$ and $PP^{INTRF}$ terms in the $PP_{dB}$ model (Eq. (10)) increases the value of $PP_{dB}$, which whittles down a large portion of the power budget $P^B_{dB}$ (Eq. (2)), leaving only a small portion of $P^B_{dB}$ available for supporting $N_\lambda$ (Eq. (2)). This results in a small value of aggregated datarate ($N_\lambda \times BR$) for a given bitrate (*BR*). Nevertheless, this ensures that the BER remains unharmed at $10^{-9}$. Therefore, if achieving the desired unharmed BER is the design goal, the terms $PP^{Fil}$, $PP^{Mod}$ and $PP^{INTRF}$ should be included in the model for $PP_{dB}$. For easy reference in the following sections of this paper, we identify such BER-optimal $PP_{dB}$ as $PP_{dB}^{BERO}$ and provide its model in Eq. (14), which includes the $PP^{Fil}$, $PP^{Mod}$ and $PP^{INTRF}$ terms.

$$PP_{dB}^{BERO} = P_L^{MR-Act} + P_L^{MR-InAct} + P_L^{WGP} + P_L^{Sp} + P_L^{INTRF} + PP^{Mod} + PP^{Fil} + PP^{PAM} + P_L^{WGB} + P_L^{C} \\ + PP^{ER} \quad (14)$$

Another way of compensating for the crosstalk-induced degradation in reliability (BER) is to use forward error correction (FEC) codes (e.g., [50], [52]). FEC codes add extra redundancy bits in every data packet to enable error detection and correction. The use of FEC codes in a photonic link can improve the BER of the link to be lower than $10^{-9}$, especially if the crosstalk inflicted BER of the link is above the typical FEC limit (e.g., $1.2\times10^{-3}$ for BCH code [50]). The use of redundant bits in FEC codes (we use the popular SECDED (72, 64) FEC [52] code in this paper) increases the packet size, and hence, the packet transfer delay and energy. Nevertheless, it does not require an increased input signal power to compensate for crosstalk-induced bit-errors. Therefore, the use of FEC codes does not whittle down the link power budget, allowing for an opportunity to support greater $N_\lambda$ and aggregated datarate in addition to achieving the desired reliability (BER). In other words, the use of FEC codes enables datarate-balanced BER in photonic links. Hence, if achieving the datarate-balanced desired BER using FEC codes is the design goal, the terms $PP^{Fil}$, $PP^{Mod}$ and $PP^{INTRF}$ need not be included in the formula for $PP_{dB}$. We identify such datarate-BER balanced $PP_{dB}$ as $PP_{dB}^{DR-BER-Bal}$ and provide its model in Eq. (15), which excludes the $PP^{Fil}$, $PP^{Mod}$ and $PP^{INTRF}$ terms.

$$PP_{dB}^{DR-BER-Bal} = P_L^{MR-Act} + P_L^{MR-InAct} + P_L^{WGP} + P_L^{Sp} + PP^{PAM} + P_L^{WGB} + P_L^{C} + PP^{ER} \quad (15)$$

These design goals (i.e., BER-optimal design versus datarate-BER balanced design) are taken into account, along with the baud-rate dependence of the detector sensitivity and the dependence of $PP_{dB}$ on MRs' spectral parameters, in our search-heuristic based optimization approach for photonic link designs, as discussed next.

## 4.2 Heuristic-Based Search for the Efficient Design of Photonic Links

Irrespective of whether the designed photonic link is BER-optimal or datarate-BER balanced, the achievable aggregated datarate (i.e., $N_\lambda \times BR$) has a cyclic dependency on the $P^B_{dB}$ and $PP_{dB}$ parameters of the link, which makes it difficult to obtain an optimal value of $N_\lambda \times BR$ for the link directly using Eqs. (1)-(15). To break this cyclic dependency and determine the optimal combination of $N_\lambda$ and $BR$ for the designed link, we employ a heuristic-based search optimization framework. The basic idea of our framework is to perform exhaustive search for the optimal combination of $N_\lambda$ and $BR$ for which the available power budget of the link ($P^B_{dB}$ in Eq. (1)) is fully utilized, while considering the factors discussed in Section 4.1 that affect the photonic link design tradeoffs.

We provide a set of baud-rate (*BaR*) and $N_\lambda$ duplets as one of the inputs to our search heuristic. We use *BaR* instead of *BR* as input because the modeling equations in Section 4.1 directly depend on *BaR*, which can be easily converted into *BR* after our search using Eq. (3). Moreover, to limit the cost and complexity of the comb-generating laser source [5], and to be consistent with the prior works on 4-PAM optical signaling [4] and [7], we limit the maximum allowable value of $N_\lambda$ to 128. Moreover, as the flit-size of a PNoC is directly proportional to the value of $N_\lambda$, and as the flit-size is usually a power-of-two value, the allowable values of $N_\lambda$ should also be power-of-two values. Because of these reasons, we choose a set Λ of all allowable values of $N_\lambda$, where Λ = {$N_\lambda$ | $N_\lambda \in$ {128, 64, 32, 16, 8, 4, 2, 1}}. Moreover, we define the set of all possible baud-rate values *R* = {*BaR* | *BaR* ∈ Q$^+$; *BaR* is in Gb/s; 10 Gb/s ≤ *BaR* ≤ 30 Gb/s; (*BaR*/0.5) ∈ N}, which has 41 elements. The individual values for Λ and *BaR* combine to make a duplet in 41×8=328 different ways. We create a set *Y* of these duplets, *Y* = {($N_{\lambda 1}$, $BaR_1$), ($N_{\lambda 1}$, $BaR_2$), …, ($N_{\lambda 8}$, $BaR_{41}$)}, and give it as an input to our search heuristic. Based on the constraint in Eq. (2), we utilize an error function *ef*($N_\lambda$, *BaR*) given in Eq. (16) to find the optimal duplet from set *Y*. For that, for each element ($N_\lambda$, *BaR*) of the set *Y*, we evaluate an error value $\epsilon$ = *ef*($N_\lambda$, *BaR*) and create a set *E* of all $\epsilon$ values. All ($N_\lambda$, *BaR*) duplets corresponding to the positive $\epsilon$ values in set *E* satisfy the constraint given in Eq. (2). But we choose the ($N_\lambda$, *BaR*) duplet corresponding to the minimum positive value $\epsilon_{min}$ from set *E* as the optimal value, because such a duplet fully utilizes the link $P^B_{dB}$.

$$ef(N_\lambda, BaR) = \{P_{dB}^B - PP_{dB} - 10\, log_{10}(N_\lambda)\} \quad (16)$$

In Eq. (16), we evaluate $PP_{dB}$ as a function of the ($N_\lambda$, *BaR*) duplet, as explained in Section 4.1. We use the search heuristic to find one ($N_\lambda$, *BaR*) duplet for every type (i.e., OOK, 4-PAM-EDAC, 4-PAM-SS, and 4-PAM-ODAC) of photonic link. We use Eqs. (14) and (15), respectively, as the models for $PP_{dB}$ in the error function (Eq. (16)) for our search of ($N_\lambda$, *BaR*) duplets. To evaluate $P_L^{WGP}$ term from Eqs. (14) and (15), we consider the maximum link length in our considered PNoCs, which is 4.5 cm for CLOS PNoC [38] and 12 cm for SWIFT PNoC [8], as provided in Table 4. The term $P_L^{SP}$ from Eqs. (14) and (15) is evaluated based on the number of splitters employed by the PNoC to power its waveguides, which differs between CLOS and SWIFT PNoCs. For example, CLOS PNoC has 56 point-to-point waveguides, and to power these waveguides, the PNoC employs 1x2, 1x7, 1x4 splitters in series [35] [38]. Therefore, the input optical power is split in 56 parts in the CLOS PNoC. Because we consider per-split loss to be 0.1 dB, the total splitter loss $P_L^{SP}$ in the CLOS PNoC is 5.6 dB, as shown in Table 4. Similarly, total splitter loss ($P_L^{SP}$) in the SWIFT PNoC is 1.2 dB which is also provided in Table 4. Note that the error-function (Eq. (16)) evaluation differs between datarate-BER balanced and BER-optimal photonic link designs, as discussed next.

### 4.2.1 Design of Datarate-BER Balanced Photonic Links

In order to design photonic links to achieve datarate-balanced BER, we do not add the modulator penalty ($PP^{Mod}$), filter penalty ($PP^{Fil}$) and signal interference penalty ($PP^{INTRF}$) terms to the total $PP_{dB}$ in Eq. (15). Because, as discussed in Section 4.1.4, $PP^{Mod}$ and $PP^{Fil}$ model crosstalk penalty when crosstalk-induced increase in BER is mitigated by increasing input optical signal power. Instead, as discussed in Section 4.1.4, we use SECDED (72, 64) FEC [52] code to counter the crosstalk-induced degradation in BER. From Section 4.1.4, using the FEC code enables the photonic links to achieve higher aggregate data rate while maintaining the BER at $10^{-9}$, thereby enabling a datarate-balanced BER value for the links. To find the optimal datarate-BER balanced ($N_\lambda$, $BaR$) duplet for a given signaling/modulation method based photonic link, we use Eq. (15) as the model for $PP_{dB}$ in the error function given in Eq. (16).

### 4.2.2 Design of BER-Optimal Photonic Links

To design BER-optimal photonic links, the modulator ($PP^{Mod}$), filter ($PP^{Fil}$) penalty terms and interference-related signal loss ($PP^{INTRF}$) are included in total $PP_{dB}$ (Eq. (14)). From Section 4.1.4, including these terms in $PP_{dB}$ in Eq. (14) results in a low aggregate datarate but the BER remains unscathed. To find the BER-optimal ($N_\lambda$, $BaR$) duplet for a given signaling/modulation method based photonic link, we use Eq. (14) as the model for $PP_{dB}$ in the error function given in Eq. (16). We repeat this exercise of finding the BER-optimal and datarate-BER balanced ($N_\lambda$, $BaR$) duplets for every CLOS and SWIFT link type (corresponding to the signaling/modulation type) for 20 nm FSR [48]. Note that our search heuristic is equitably applicable to the OOK, 4-PAM-SS, 4-PAM-EDAC and 4-PAM-ODAC links. However, the optimal ($N_\lambda$, $BaR$) duplets would differ for different link types, as the values of $PB_{dB}$ and other design parameters differ for different link types.

## 4.3 Results of Optimal Designs of Photonic Links Using Heuristic-Based Search

In this section, we present our obtained BER-optimal and datarate-BER balanced ($N_\lambda$, $BaR$) duplets for different variants of the CLOS and SWIFT links (i.e., 4.5 cm long links for CLOS PNoC [38] and 12 cm long links for SWIFT PNoC [8], with various modulation methods) for an FSR value of 20 nm. We also report our evaluated aggregated datarate ($N_\lambda \times BR$) and $PB_{dB}$ values for different variants of CLOS and SWIFT links. To evaluate aggregated datarate, we use Eq. (3) to convert the $BaR$ values found through our search heuristic in the corresponding $BR$ values.

### 4.3.1 Results for Datarate-BER Balanced Links

Table 5 gives optimal $N_\lambda$, bitrate ($BR$) (evaluated from $BaR$), aggregated datarate ($N_\lambda \times BR$), and $PB_{dB}$ values for different datarate-BER balanced variants of CLOS and SWIFT links. It also gives $PP_{dB}$ + 10log($N_\lambda$) values for our considered link variants. For brevity, we do not provide $PP_{dB}$ values, but these values can be easily derived from $PP_{dB}$ + 10log($N_\lambda$) values as $N_\lambda$ values are already provided in Table 5.

From Table 5, the aggregated datarate values for various SWIFT links are in general lower than the aggregated datarate values for various CLOS links. This is because links in SWIFT have greater $P_L^{Sp}$, $P_L^C$, and $P_L^{WGP}$ values in Eq. (15) than the CLOS links (Table 4), which results in larger $PP_{dB}$ values for the SWIFT links. As a result, a relatively smaller portion $PB_{dB}$ is available in Eq. (16) for the SWIFT links to support the aggregated datarate, resulting in smaller $N_\lambda$ and $N_\lambda \times BR$ (aggregated datarate) values for the SWIFT links. Further, it is interesting to note that the 4PAM-EDAC, 4PAM-ODAC, and OOK based CLOS links can achieve >1,000Gb/s (>1Tb/s) aggregated datarate. This outcome is in strong agreement with the performance analysis done for photonic links in prior works [34] and [48].

However, none of the SWIFT links can achieve >1Tb/s aggregated datarate, which corroborates the observation made in [46] that to achieve terascale aggregate data rates in photonic links the losses and power penalties in the links must be minimized. As per our analysis, the CLOS links have significantly low losses and penalties compared to the SWIFT links, and as a result, the CLOS links can achieve >1Tb/s datarate, whereas the SWIFT links cannot.

In addition, Table 5 also lists BER values for various CLOS and SWIFT links evaluated when SECDED coding was not used. These values give insights into how the crosstalk noise present in the links impacts BER. To evaluate the BER for a link, we evaluated the worst-case $P_{Xtalk}$ from Eq. (7) across all the filters in the receiving filter MR bank when considering the $P_{NRZ}$ in Eq. (7) to be the signal power reaching the worst-case filter MR after accounting for all the losses and penalties encountered in the link based on Eq. (15). From there, we evaluated the signal-to-noise ratio (SNR) to be $P_{NRZ}/P_{Xtalk}$. Based on the mathematical models and equations of BER provided for guided propagation in [20], [73], we formulate a relation between BER and SNR in Eq. (17). M in Eq. (17) is defined in Table 3. From Eq. (17), we formulated the BER equations for OOK (M = 2) and 4-PAM (M = 4) signals, which are provided in Eqs. (18) and (19), respectively. We have used the evaluated SNR value in Eq. (18) (from [12]) for OOK links and Eq. (19) (from [12]) for 4-PAM links to determine the corresponding BER values, which are reported in Table 5.

From these BER values (shown in Table 5), it is evident that all datarate-BER balanced CLOS and SWIFT links achieve the BER values that are lower than $1.74 \times 10^{-3}$, which is a threshold BER value for the SECDED (64, 72) coding to achieve error-free transmission of data packets of size 512 bits (we consider 512-bits long packets in our system-level evaluation in the next section). We obtain this threshold value through the following reasoning: the threshold BER value in a SECDED (64, 72) coded data packet should not incur more than 1-bit of error, as only 1-bit can be corrected for, to achieve error-free transmission of the SECDED (64, 72) coded data packet. Since a 512-bit original data packet gets converted into a 576-bit data packet after it is encoded with the SECDED (64, 72) code, up to only 1-bit in the 576-bit data packet is allowed to be erroneous. Therefore, the threshold BER for this case becomes, $1/576 = 1.74 \times 10^{-3}$. Thus, all the CLOS and SWIFT links in Table 5 are capable of achieving error-free data transmission using SECDED (64,72) coding, as all links in Table 5 achieve BER of lower than $1.74 \times 10^{-3}$, which ensures that all possible bit errors in the 512-bit data packets transmitted over these links can be corrected using the SECDED (64,72) coding.

Moreover, comparing the aggregated datarate values for the links with different modulation methods, it is evident that 4PAM-EDAC and 4PAM-ODAC links in general achieve greater aggregated datarate compared to OOK links. This is because, compared to the OOK links that can have only one bit transferred per signal symbol, the 4PAM-EDAC and 4PAM-ODAC links can achieve greater *BR* due to their ability to transfer 2-bits per signal symbol. As a result, the 4PAM-EDAC and 4PAM-ODAC links achieve greater values of aggregated datarate ($N_\lambda \times BR$), despite achieving the same $N_\lambda$ values as achieved by the OOK links. On the other hand, comparing the different types of 4PAM links with one another, it is evident that: *(i)* the 4PAM-SS links achieve lower aggregated datarate values than the 4PAM-EDAC and 4PAM-ODAC links; and *(ii)* the 4PAM-ODAC links achieve aggregated datarate values that are higher than the 4PAM-SS links but lower than the 4PAM-EDAC links. This is because: *(i)* the 4PAM-SS links have the largest $PP_{dB}$ values due to their higher MR through losses caused due to 2× more MR modulators required for them (Table 1, Fig. 2), which results in the lowest $N_\lambda$ values for them, compared to the 4PAM-EDAC and 4PAM-ODAC links; and *(ii)* the 4PAM-ODAC links have greater $PP^{ER}$ value compared to the 4PAM-EDAC links (Table 4), which results in greater $PP_{dB}$ values for the 4PAM-ODAC links, yielding lower values of available $P^B_{dB}$ that support lower *BR* values (and hence, lower $N_\lambda \times BR$) for the 4PAM-ODAC links.

$$\text{BER} = \frac{2(M-1) - \log_2 M}{M \times \log_2 M} \text{erfc}\left(\frac{\sqrt{\text{SNR}}}{(M-1)\sqrt{2}}\right) \qquad (17)$$

$$\text{BER}_{\text{OOK}} = \frac{1}{2} \text{erfc}\left(\frac{\sqrt{\text{SNR}}}{\sqrt{2}}\right) \qquad (18)$$

$$\text{BER}_{4-\text{PAM}} = \frac{1}{2} \text{erfc}\left(\frac{\sqrt{\text{SNR}}}{3\sqrt{2}}\right) \qquad (19)$$

**Table 5: Optimal $N_\lambda$, bitrate (*BR*), aggregate datarate ($N_\lambda \times BR$), power budget ($P^B_{dB}$), $PP_{dB}$ + 10log($N_\lambda$), detector sensitivity (*S*), and optical laser power (= $PP_{dB}$ + *S*) values for different datarate-BER balanced variants of CLOS and SWIFT links evaluated for different extinction ratios of OOK and 4-PAM modulation schemes**

| Variants | Extinction Ratio (dB) | $P^B_{dB}$ (dB) | Detector Sensitivity (*S*) | Optimal $N_\lambda$ | BR (Gb/s) | $N_\lambda \times BR$ (Gb/s) | $PP_{dB}$ +10log($N_\lambda$) (dB) | Optical Laser Power | *BER Without SECDED Coding* |
|---|---|---|---|---|---|---|---|---|---|
| Various CLOS Links for FSR = 20nm [48] | | | | | | | | | |
| OOK | 5 | 38.60 | -18.6 dBm | 64 | 17 | 1,088 | 38.29 | 19.69 dBm | $3.39 \times 10^{-5}$ |
| OOK | 9 | 37.80 | -17.80 dBm | 64 | 18 | 1,152 | 37.31 | 19.51 dBm | $2.9 \times 10^{-5}$ |
| OOK | 12 | 37.1 | -17.1 dBm | 64 | 19 | 1,216 | 36.31 | 19.21 dBm | $2.6 \times 10^{-5}$ |
| 4-PAM-SS | 5 | 42.50 | -22.5 dBm | 32 | 20 | 640 | 41.80 | 19.3 dBm | $12 \times 10^{-4}$ |
| 4-PAM-SS | 9 | 41.00 | -21 dBm | 32 | 27 | 864 | 40.8 | 19.8 dBm | $7.9 \times 10^{-4}$ |
| 4-PAM-SS | 12 | 40.35 | -20.35 dBm | 32 | 30 | 960 | 39.8 | 19.45 dBm | $7.5 \times 10^{-4}$ |
| 4-PAM-EDAC | 5 | 40.35 | -20.35 dBm | 64 | 30 | 1,920 | 38.00 | 17.65 dBm | $8.83 \times 10^{-5}$ |
| 4-PAM-EDAC | 9 | 37.9 | -17.9 dBm | 64 | 35 | 2,240 | 37.00 | 19.1 dBm | $8 \times 10^{-5}$ |
| 4-PAM-EDAC | 12 | 36.1 | -16.1 dBm | 64 | 40 | 2,560 | 36.00 | 19.9 dBm | $6.2 \times 10^{-5}$ |
| 4-PAM-ODAC | 2 | 42.50 | -22.5 dBm | 64 | 20 | 1,280 | 42.00 | 19.5 dBm | $9.7 \times 10^{-4}$ |
| 4-PAM-ODAC | 6 | 38.5 | -18.5 dBm | 64 | 33 | 2,112 | 38.31 | 19.81 dBm | $8.3 \times 10^{-5}$ |
| 4-PAM-ODAC | 9 | 37.9 | -17.9 dBm | 64 | 35 | 2,240 | 37.41 | 19.51 dBm | $8 \times 10^{-5}$ |
| Various SWIFT links for FSR = 20nm [48] | | | | | | | | | |
| OOK | 5 | 38.60 | -18.6 dBm | 32 | 17 | 544 | 38.06 | 19.46 dBm | $8.02 \times 10^{-5}$ |
| OOK | 9 | 37.80 | -17.80 dBm | 32 | 18 | 576 | 37.1 | 19.30 dBm | $7.8 \times 10^{-5}$ |
| OOK | 12 | 37.1 | -17.1 dBm | 32 | 19 | 608 | 36.1 | 19 dBm | $7.1 \times 10^{-5}$ |
| 4PAM-SS | 5 | 42.10 | -22.1 dBm | 16 | 22 | 352 | 40.85 | 18.75 dBm | $1.4 \times 10^{-3}$ |
| 4PAM-SS | 9 | 40.35 | -20.35 dBm | 16 | 30 | 480 | 39.9 | 19.55 dBm | $1 \times 10^{-3}$ |
| 4PAM-SS | 12 | 39.1 | -19.1 dBm | 16 | 32 | 512 | 38.9 | 19.8 dBm | $9 \times 10^{-4}$ |
| 4PAM-EDAC | 5 | 41.00 | -21 dBm | 32 | 27 | 864 | 38.16 | 17.16 dBm | $8.17 \times 10^{-4}$ |
| 4PAM-EDAC | 9 | 37.9 | -17.9 dBm | 32 | 35 | 1,120 | 37.2 | 19.3 dBm | $7 \times 10^{-4}$ |
| 4PAM-EDAC | 12 | 41 | - 21 dBm | 64 | 27 | 1,728 | 40.7 | 19.7 dBm | $3.5 \times 10^{-4}$ |
| 4PAM-ODAC | 2 | 42.30 | -22.3 dBm | 32 | 21 | 672 | 42.14 | 19.84 dBm | $8.49 \times 10^{-4}$ |
| 4PAM-ODAC | 6 | 38.5 | -18.5 dBm | 32 | 33 | 1,056 | 38.41 | 19.91 dBm | $7.3 \times 10^{-4}$ |
| 4PAM-ODAC | 9 | 42.5 | -22.5 dBm | 64 | 20 | 1,280 | 42.31 | 19.81 dBm | $7.6 \times 10^{-4}$ |

Table 5 also provides optimal $N_\lambda$, total power penalty ($PP_{dB}$ + 10log($N_\lambda$), aggregate datarate and optical laser power consumption of OOK and PAM4 based datarate-BER balanced variants of CLOS and SWIFT links, corresponding to different extinction ratios. An increase in the extinction ratio reduces the extinction ratio penalty ($PP^{ER}$) and filter penalty (Eq. (6)), which in turn reduces the total penalty in the link ($PP_{dB}$). This reduction in total power penalty creates more room in the link power budget ($P^B_{dB}$) to be leveraged to achieve larger $N_\lambda$ and/or increased bit-rate (BR), and hence, achieve larger aggregate datarate (i.e., $N_\lambda \times$ BR) for the link. However, the laser power consumption does not improve noticeably with the increase in extinction ratio, as evident from Tables 5 and 6. This is because since the laser power consumption is given as $PP_{dB}$ + S, the laser power reduces only if the combined effect of the reduction in $PP_{dB}$ and/or increase in S reduces $PP_{dB}$ + S. More specifically, from Table 5, with the increase in the extinction ratio from 5 dB to 12 dB, the aggregate datarates of the OOK based CLOS and SWIFT links increase from 1.08 Tb/s to 1.22 Tb/s and from 544 Tb/s to 608 Tb/s respectively. This is because for the OOK based CLOS links the BR increases from 17 Gb/s to 19 Gb/s at the unchanged $N_\lambda$ of 64, and for the OOK based SWIFT links the BR increases from 17 Gb/s to 19 Gb/s at the unchanged $N_\lambda$ of 32. In terms of laser power consumption, OOK based CLOS and SWIFT links experience reduction in laser power consumption from 19.7 dBm to 19.21 dBm and 19.5 dBm to 19 dBm respectively with increase in extinction ratio from 5 dB to 12 dB. This is because OOK based CLOS and SWIFT links achieve reduced ($PP_{dB}$ +S) because of combined effects of reduction in $PP_{dB}$ due to increase in extinction ratio and increase in S due to increase in BR. Similarly, with the increase in extinction ratio from 5 dB to 12 dB, the aggregate datarates of the 4-PAM-EDAC based CLOS and SWIFT links increase from 1.92 Tb/s to 2.6 Tb/s and 864 Gb/s to 1.73 Tb/s respectively. This is because for 4-PAM-EDAC based CLOS links, the BR increases from 30 Gb/s to 40 Gb/s at the unchanged $N_\lambda$ of 64, and for 4-PAM-EDAC based SWIFT links, the $N_\lambda$ increases from 32 to 64. In terms of laser power consumption, 4-PAM-EDAC based CLOS and SWIFT links experience an increase in laser power consumption from 17.65 dBm to 19.9 dBm and 17.16 dBm to 19.7 dBm respectively, with increase in extinction ratio from 5 dB to 12 dB. This is because for 4-PAM EDAC based CLOS links, the decrease in $PP_{dB}$ due to the increase in extinction ratio is offset by the larger increase in S due to the increase in BR from 30 Gb/s to 40 Gb/s. In contrast, for the 4-PAM-EDAC based SWIFT links, the increase in $N_\lambda$ from 32 to 64 increases $PP_{dB}$, which in turn increases the laser power consumption of the link. Similarly, with the increase in extinction ratio from 2 dB to 9 dB, the aggregate datarates of the 4-PAM-ODAC based CLOS and SWIFT links increase from 1.3 Tb/s to 2.24 Tb/s and 672 Gb/s to 1.3 Tb/s respectively This is because for 4-PAM-ODAC based CLOS links, the BR increases from 20 Gb/s to 35 Gb/s with the unchanged $N_\lambda$ of 64 and for 4-PAM ODAC based SWIFT links, the $N_\lambda$ increases from 32 to 64 with decrease in BR from 21 Gb/s to 20 Gb/s. In terms of laser power consumption, 4-PAM-ODAC based CLOS links experience an increase in laser power consumption from 19.5 dBm to 19.51 dBm similar to 4-PAM-EDAC based CLOS links. On the other hand, 4-PAM ODAC based SWIFT links experience slight reduction in laser power consumption from 19.84 dBm to 19.81 dBm with increase in extinction ratio. Also, with increase in extinction ratio from 5 dB to 12 dB, the aggregate datarates of 4-PAM-SS based CLOS and SWIFT links increase from 640 Gb/s to 960 Gb/s and 352 Gb/s to 512 Gb/s respectively. In terms of laser power consumption, similar to 4-PAM-EDAC based links, 4-PAMSS based CLOS and SWIFT links experience increase in laser power consumption from 19.3 dBm to 19.45 dBm and 18.75 dBm to 19.8 dBm respectively with increase in extinction ratio.

*In summary*, the datarate-BER balanced 4PAM-EDAC links achieve the highest datarate across the CLOS and SWIFT link types. However, it is not clear from these datarate results if the 4PAM-EDAC links can be more energy-efficient than the other types of links. To determine whether the higher overhead of the modulator driver energy for the 4PAM-EDAC links (Table 1 and Table 2) can offset their highest datarate related benefits to yield lower energy-efficiency for them, compared to the OOK and other 4PAM links, we performed a system-level analysis with real-world benchmark applications, the details of which are discussed in Section 5.

### 4.3.2 Results for BER-Optimal Links

Table 6 shows the optimal $N_\lambda$, bitrate (*BR*) (evaluated from *BaR*), aggregated datarate ($N_\lambda \times BR$), $P^B_{dB}$ values, and $PP_{dB} + 10\log(N_\lambda)$ values for different BER-optimal variants of CLOS and SWIFT links. Similar to the results for the datarate-BER balanced links presented in Table 5, the results presented in Table 6 also lead to the following observations: *(i)* The CLOS links achieve higher datarate than the SWIFT links across all evaluated modulation types, due to the lower $P_L^{Sp}$, $P_L^C$, and $P_L^{WGP}$ values for the CLOS links than the SWIFT links. *(ii)* The 4PAM-EDAC links achieve the highest datarate values across the CLOS and SWIFT types, because 4PAM-EDAC links have the lowest $PP_{dB}$ values, compared to the OOK and other 4PAM links. *(iii)* The 4PAM-SS links achieve the lowest datarate values because they have the highest $PP_{dB}$ values due to the non-zero $PP^{INTRF}$ for them (Table 4) and their higher MR through losses caused due to 2× more MR modulators required for them (Table 1, Fig. 2).

Table 6 also provides optimal $N_\lambda$, total loss, aggregate datarate and optical laser power consumption of OOK and PAM4 based BER-optimal variants of CLOS and SWIFT links corresponding to different extinction ratios. As we can infer from Table 6, with increase in extinction ratio from 5 dB to 12 dB, the aggregate data rates of OOK based CLOS and SWIFT links increase from 864 Gb/s to 1.02 Tb/s and 512 Gb/s to 672 Gb/s respectively. This is because for OOK based CLOS links, the BR increases from 27 Gb/s to 32 Gb/s at the unchanged $N_\lambda$ of 32 and for OOK based SWIFT links, the BR increases from 16 Gb/s to 21 Gb/s at the unchanged $N_\lambda$ of 32. In terms of laser power consumption, with increase in extinction ratio from 5 dB to 12 dB, laser power consumption of OOK based CLOS links reduces from 19.9 dBm to 19.7dBm whereas laser power consumption of OOK based SWIFT links increases from 19.66 dBm to 19.8 dBm. This is because OOK based CLOS links achieve reduced ($PP_{dB}$ +S) because of combined effects of reduction in $PP_{dB}$ due to increase in extinction ratio and increase in S due to increase in BR which in turn results in reduced laser power consumption. On the other hand, for OOK based SWIFT links, the decrease in $PP_{dB}$ due to increase in extinction ratio is offset by larger values of S with increase in BR from 16 Gb/s to 21 Gb/s. Similarly, with increase in extinction ratio from 2 dB to 9 dB, the aggregate data rates of 4-PAM-ODAC based CLOS and SWIFT links increase from 768 Gb/s to 1.6 Tb/s and 352 Gb/s to 960 Gb/s respectively. This is because for 4-PAM-ODAC based CLOS links, the BR increases from 24 Gb/s to 50 Gb/s at the unchanged $N_\lambda$ of 32 and for 4-PAM-ODAC based SWIFT links, the BR increases from 22 Gb/s to 30 Gb/s with increase in $N_\lambda$ from 16 to 32. In terms of laser power consumption, with increase in extinction ratio from 2 dB to 9 dB, laser power consumption of 4-PAM-ODAC based CLOS links reduces from 19.6 dBm to 19.5 dBm. On the other hand, laser power consumption of 4-PAM-ODAC based SWIFT links increases from 19.65 dBm to 19.8 dBm. This is because for 4-PAM-ODAC based CLOS links, combined effect of reduction in $PP_{dB}$ due increase in extinction ratio and increase in S due to increase in BR from 24 Gb/s to 50 Gb/s reduces laser power consumption of the link. For 4-PAM-ODAC based SWIFT links, increase in $PP_{dB}$ due to increase in $N_\lambda$ from 16 to 32 increases laser power consumption of the link. Similarly, For 4-PAM EDAC links, with increase in extinction ratio from 5 dB to 9 dB, the aggregate data rates increase from 1.02 Tb/s to 1.5 Tb/s for CLOS links and 512 Gb/s to 736 Gb/s for SWIFT links. This is because for 4-PAM-EDAC based CLOS links, the BR increases from 32 Gb/s to 48 Gb/s at the unchanged $N_\lambda$ of 32 and for 4-PAM-EDAC based SWIFT links, the BR increases from 32 Gb/s to 46 Gb/s at the unchanged $N_\lambda$ of 16. In terms of laser power consumption, with increase in extinction ratio from 5 dB to 12 dB, laser power consumption of 4-PAM-EDAC based CLOS links increase from 18.13 dBm to 19.13 dBm and laser power consumption of 4-PAM-EDAC based SWIFT links reduces from 19.7 dBm to 19.6 dBm. This is because for 4-PAM-EDAC based CLOS links, the reduction in $PP_{dB}$ due to increase in extinction ratio is nullified by larger increase in S due to increase in BR from 32 Gb/s to 48 Gb/s which in turn increases the laser power consumption of the link. In contrast, for 4-PAM-EDAC based SWIFT links, the combined effect of reduction in $PP_{dB}$ and increase in S results in reduced laser power consumption of the link. Also, For 4-PAM-SS based links, with increase in extinction ratio from 5 dB to 12 dB, the aggregate data rates increase from 352 Gb/s to 608 Gb/s for CLOS links and 160 Gb/s to 384 Gb/s for SWIFT links. This is because for 4-PAM-SS based CLOS links, the BR increases from 22 Gb/s to 38 Gb/s at unchanged $N_\lambda$ of 16 and for 4-PAM-SS based SWIFT links, the BR increases from 20 Gb/s to 24 Gb/s with increase in $N_\lambda$ from 8 to 16. In terms of laser power consumption, similar to 4-PAM-ODAC links, with increase in extinction ratio from 5 dB to 12 dB, laser power consumption of 4-PAM-SS based CLOS links decreases from 19.93 dBm to 19.1 dBm because of the combined effect of reduction in $PP_{dB}$ due to increase in extinction ratio and increase in S due to increase in BR from 22 Gb/s to 38 Gb/s. On the other hand, laser power consumption of 4-PAM-SS based SWIFT links increases from 17.86 dBm to 19.3 dBm due to increase in $PP_{dB}$ since $N_\lambda$ increases from 8 to 16.

In addition, it can be observed that the OOK links achieve higher datarate values than the 4PAM-ODAC and 4PAM-SS links. This is because the inclusion of $PP^{Fil}$, $PP^{Mod}$ and $PP^{INTRF}$ terms in Eq. (14) increases the $PP_{dB}$ values for the 4PAM-ODAC and 4PAM-SS links to be greater than the $PP_{dB}$ values for the OOK links, which results in higher values of available $P^{B}_{dB}$ for the OOK links, leading to higher aggregated datarate ($N_\lambda \times BR$) for the OOK links. Due to the inclusion of $PP^{Fil}$, $PP^{Mod}$ and $PP^{INTRF}$ terms in Eq. (14), only the 4PAM-EDAC links among all the three different 4PAM link types achieve greater datarate than the OOK links. However, it is not clear if these datarate benefits can allow 4PAM-EDAC to achieve better energy-efficiency than the OOK links. This is because the greater number of hardware components required for realizing the 4PAM-EDAC links (see the #serialization units, #deserialization units, and #transimpedance op-amps in Table 1) can offset their datarate benefits to render them with lower energy-efficiency, compared to the OOK links. To investigate this possibility, we performed a system-level (PNoC-level) analysis with real-world benchmark applications, the details of which are discussed in Section 5.

**4.3.3   Datarate-BER Balanced vs BER-Optimal Links**

From Table 5 and Table 6, it is evident that the datarate-BER balanced links in general achieve higher aggregated datarate than the BER-optimal links. This is because for the BER-optimal links, due to the inclusion of the terms $PP^{Mod}$, $PP^{Fil}$ and $PP^{INTRF}$ for the evaluation of $PP_{dB}$, more of the provisioned optical power is utilized for ensuring the target reliability in terms of the target BER ($10^{-9}$ in this paper). As a result, a relatively small amount of the total provisioned optical power remains available to support the aggregated datarate for the BER-optimal links, leading to relatively lower aggregate datarate. In contrast, for the datarate-BER balanced links, the exclusion of the terms $PP^{Mod}$, $PP^{Fil}$ and $PP^{INTRF}$ from the $PP_{dB}$ formula keeps a relatively large amount of the total provisioned optical power available in the links, which supports relatively large values of aggregated datarate for the datarate-BER balanced links. Despite achieving large datarate values, the datarate-BER balanced links may still achieve lower average performance and energy-efficiency compared to the BER-optimal links, especially when they are utilized in a PNoC. This is because, the datarate-BER balanced links in general utilize redundant bits of the SECDED (64, 72) coding in every data packet that traverses the PNoC, which can result in a relatively higher average packet latency and per-packet dynamic energy consumption, ultimately leading to a lower value of the average throughput and energy-efficiency in the PNoC. To investigate this hypothesis, we performed a system-level (PNoC-level) analysis with real-world benchmark applications, the details of which are discussed next, in Section 5.

Table 6: Optimal $N_\lambda$, bitrate ($BR$), aggregated datarate ($N_\lambda \times BR$), power budget ($P^B{}_{dB}$), $PP_{dB}$ + 10log($N_\lambda$), detector sensitivity ($S$), and optical laser power (= $PP_{dB}$ + $S$) values for different BER optimal variants of CLOS and SWIFT links evaluated for different extinction ratios of OOK and 4-PAM modulation schemes

| Variants | Extinction Ratio (dB) | $P^B{}_{dB}$ (dB) | Detector Sensitivity (S) | Optimal $N_\lambda$ | BR (Gb/s) | $N_\lambda \times BR$ (Gb/s) | $PP_{dB}$ +10log($N_\lambda$) (dB) | Optical Laser Power |
|---|---|---|---|---|---|---|---|---|
| Various CLOS links for FSR = 20nm [48] | | | | | | | | |
| OOK | 5 | 30.10 | -10.1 dBm | 32 | 27 | 864 | 30.00 | 19.9 dBm |
| OOK | 9 | 28.2 | -8.2 dBm | 32 | 30 | 960 | 27.63 | 19.43 dBm |
| OOK | 12 | 26.6 | -6.6 dBm | 32 | 32 | 1,024 | 26.3 | 19.7 dBm |
| 4PAM-SS | 5 | 42.10 | -22.1 dBm | 16 | 22 | 352 | 42.03 | 19.93 dBm |
| 4PAM-SS | 9 | 37.9 | -17.9 dBm | 16 | 35 | 560 | 37.7 | 19.8 dBm |
| 4PAM-SS | 12 | 37.1 | -17.1 dBm | 16 | 38 | 608 | 36.2 | 19.1 dBm |
| 4PAM-EDAC | 5 | 39.10 | -19.1 dBm | 32 | 32 | 1,024 | 37.23 | 18.13 dBm |
| 4PAM-EDAC | 9 | 33.4 | -13.4 dBm | 32 | 46 | 1,472 | 32.93 | 19.53 dBm |
| 4PAM-EDAC | 12 | 32.3 | -12.3 dBm | 32 | 48 | 1,536 | 31.43 | 19.13 dBm |
| 4PAM-ODAC | 2 | 41.70 | -21.7 dBm | 32 | 24 | 768 | 41.30 | 19.6 dBm |
| 4PAM-ODAC | 6 | 32.3 | -12.3 dBm | 32 | 48 | 1,536 | 32.1 | 19.8 dBm |
| 4PAM-ODAC | 9 | 31.5 | -11.5 dBm | 32 | 50 | 1,600 | 31 | 19.5 dBm |
| Various SWIFT links for FSR = 20nm [48] | | | | | | | | |
| OOK | 5 | 39.10 | -19.1 dBm | 32 | 16 | 512 | 38.76 | 19.66 dBm |
| OOK | 9 | 37.1 | -17.1 dBm | 32 | 19 | 608 | 36.4 | 19.3 dBm |
| OOK | 12 | 35.3 | -15.3 dBm | 32 | 21 | 672 | 35.1 | 19.8 dBm |
| 4PAM-SS | 5 | 42.50 | -22.5 dBm | 8 | 20 | 160 | 40.36 | 17.86 dBm |
| 4PAM-SS | 9 | 37.1 | -17.1 dBm | 8 | 38 | 304 | 36.1 | 19 dBm |
| 4PAM-SS | 12 | 41.7 | -21.7 dBm | 16 | 24 | 384 | 41 | 19.3 dBm |
| 4PAM-EDAC | 5 | 39.10 | -19.1 dBm | 16 | 32 | 512 | 38.80 | 19.7 dBm |
| 4PAM-EDAC | 9 | 35.3 | -15.3 dBm | 16 | 42 | 672 | 34.5 | 19.2 dBm |
| 4PAM-EDAC | 12 | 33.4 | -13.4 dBm | 16 | 46 | 736 | 33 | 19.6 dBm |
| 4PAM-ODAC | 2 | 42.10 | -22.1 dBm | 16 | 22 | 352 | 41.75 | 19.65 dBm |
| 4PAM-ODAC | 6 | 41.7 | -21.7 dBm | 32 | 24 | 768 | 41.31 | 19.61 dBm |
| 4PAM-ODAC | 9 | 40.35 | -20.35 dBm | 32 | 30 | 960 | 40.15 | 19.8 dBm |

## 5 System level evaluation

### 5.1 Evaluation Setup and Methodology

For evaluating optimized link-level variants based on OOK and several 4-PAM modulation schemes at system-level, we have considered two separate PNoC architectures: CLOS PNoC [38] and SWIFT PNoC [8]. We particularly selected the photonic crossbar based, high-radix SWIFT PNoC architecture [8] for this system-level analysis, because SWIFT PNoC has been shown in [8] to provide significantly better throughput and energy-efficiency compared to the other classic high-radix PNoC architectures, such as [72]. In addition, to evaluate another high-radix PNoC architecture that is distinct from the photonic crossbar based SWIFT PNoC, we also selected the 8-ary 3-stage CLOS PNoC architecture from [38] that employs WDM based point-to-point photonic links. We preferred high-radix PNoC architectures to more classic, low-radix architectures such as [70] and [71], as prior works [72], [8], and [84] have shown that high-radix PNoC architectures are extremely promising architectures to meet future on-chip bandwidth demands. These PNoC architectures were evaluated for the following modulation schemes: OOK, 4-PAM-SS, 4-PAM-EDAC, and 4-PAM-ODAC.

For CLOS-PNoC [38] shown in Fig. 7(a), we have considered an 8-ary 3-stage topology for a 256 x86-core system with 8 clusters, 8 tiles in each cluster, and 4 cores in each tile. The 4 cores of each tile connect with one another via a concentrator. The 8 concentrators corresponding to the 8 tiles in a cluster communicate with one another via an electrical router. The electrical router is a simple 8×8 router, with each concentrator connected to the router using one of its ports. The concentrators and electrical routers are not shown in Fig. 7(a). Each router is associated with a photonic transmitter and receiver block (Fig. 7(a)), and the electrical-optical-electrical conversion happens at the photonic transmitter-receiver block. For inter-cluster communication, point to point photonic waveguides are supported by the photonic transmitter-receiver blocks, with forward or backward propagating wavelengths depending upon the physical location of the source and destination clusters. All the clusters are connected together using total 56 waveguides (WGs). The PNoC uses two laser sources to enable bi-directional communication.

For SWIFT PNoC [8] shown in Fig. 7(b), we have again considered a 256 x86-core system. Every 4-core cluster is considered a node here and communication within a node occurs through a 5×5 electrical router. Four ports of the router connect the processing cores to the router and the fifth port of the router is connected to a gateway interface (GI) which facilitates transfers between the electrical and photonic layers. The routers use round-robin arbitration to facilitate communication between cores and the GIs. Each GI connects four nodes. The architecture utilizes a photonic crossbar topology with eight waveguide groups and four Multiple Writer Multiple Reader (MWMR) WGs per group. A broadband off-chip laser with a laser power controller is used to power the WGs.

We consider the two-layer, 3D chip organization from [8] and [83] for each of these PNoC architectures. The bottom CMOS layer contains processing cores, caches, and electrical interconnects. The silicon-photonic top layer contains photonic transmitter-receiver (Tx-Rx) blocks (for O/E and E/O conversions), as well as photonic devices and circuits that constitute a PNoC. Through-silicon vias (TSVs) are used to vertically connect the bottom layer with the top layer at every photonic Tx-Rx block. The parameters used for modeling the 3D organizations of our considered PNoCs are given in Table 7. The $N_\lambda$ and BR values in Table 7 can be taken from Table 5 and 6, depending on the utilized modulation scheme and target design goal (datarate-BER balanced or BER-optimal design).

For evaluating the impact of various signaling methods on these architectures' performance and efficiency, we performed a benchmark-driven simulation-based analysis using Gem5 full-system simulation [57] and an enhanced cycle-accurate PNoC simulator that extends the Noxim simulator [58]. For Gem5 simulations, we assumed 32KB direct mapped L1 and 128 KB direct mapped L2 caches (MOESI coherency for L2) per core and a main memory of 32 GB DDR4 RAM. Tables 5 and 6 show the number of wavelengths ($N_\lambda$) and the maximum datarate for which the simulations were run for CLOS PNoC and SWIFT PNoC. These $N_\lambda$ and datarate values were utilized to model the links in different variants of the CLOS and SWIFT PNoCs that correspond to various signaling schemes and datarate-BER balanced and BER-optimal designs. The energy and power value considerations from Tables 1 and 2 were also incorporated into our simulations. The performance was evaluated at a 22nm CMOS node. The floorplan and the number of WGs were kept constant across all variants of a particular PNoC architecture, with only the link configuration parameters (e.g., $N_\lambda$, datarate, number of hardware instances from Table 1 and Table 2 that depend on $N_\lambda$) changing across the variants. PARSEC benchmark applications [15] were used to generate real-world traffic traces. The traces were generated using GEM5 full-system simulation and these traces were fed into our cycle accurate PNoC-simulator. In GEM5 simulations, the warm-up period was set as 100M cycles and the traces were captured for the subsequent 1B instructions. The simulations were used to evaluate average latency, energy-per-bit (EPB) and a breakdown of total power dissipation. Electrical energy consumption by routers and GIs was determined using the DSENT tool [17]. To obtain the laser power consumption, the total required optical power in the CLOS and SWIFT PNoC architectures were evaluated based on the $P^B_{dB}$ and $PP_{dB}$ values from Tables 5 and 6 for different variants, and then 15% wall-plug efficiency [55] was assumed to convert these optical power values into the corresponding electrical laser power values. The energy and power value considerations from Tables 1 and 2 were also incorporated into our simulations. The performance was evaluated at a 22nm CMOS node. The floorplan and the number of WGs were kept constant across all variants of a particular PNoC architecture, with only the link configuration parameters (e.g., $N_\lambda$, datarate, number of hardware instances from Table 1 and Table 2 that depend on $N_\lambda$) changing across the variants.

To implement SECDED encoding for the datarate-BER balanced variants of the CLOS and SWIFT PNoC architectures, we employed a lookup table-based approach where each input 512-bit packet is encoded with the SECDED scheme via byte-level lookup tables. In other words, every Byte of the input packet gets encoded through a separate and parallelly operating lookup table. Because of this parallelism in encoding, this encoding incurs only a one cycle delay. SECDED decoding is also handled using byte-wise parallelly operating lookup tables. However, the one cycle delay of only the decoding phase comes in the critical latency path in the PNoCs, as the encoding delay can be hidden by overlapping the encoding operation with the arbitration and receiver selection phases in the PNoC. We considered the area overhead of the SRAM-implemented lookup tables (at the 22 nm node) in the encoding and decoding units, and estimated it to be 1142 μm² each. Each GI in the PNoC should have one encoding unit and one decoding unit, therefore, each GI would have 2284 μm² of area overhead. In addition, each encoding or decoding event for a 512-bit packet is estimated to consume 0.1 pJ energy. The area estimates were obtained using logic synthesis analysis. The energy and delay values were evaluated using CACTI-P [56] and are accounted for in our system-level analysis.

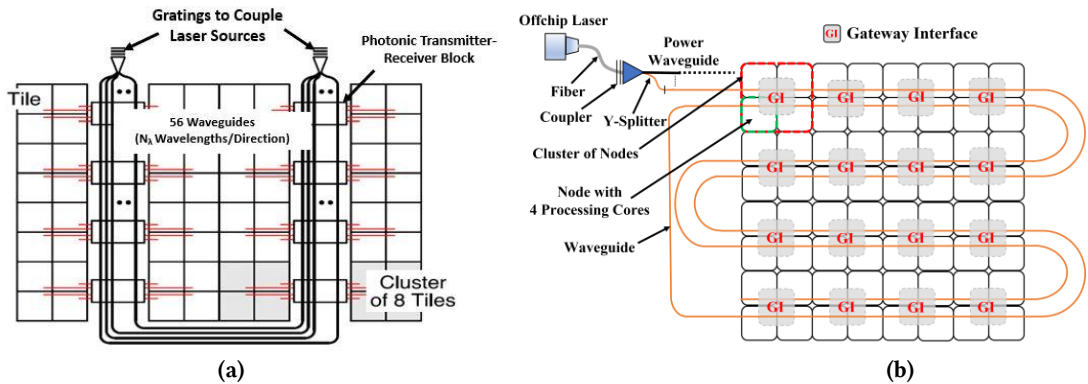

Fig. 7: Schematics of (a) 8-ary 3-stage CLOS-PNoC architecture [38] and (b) SWIFT PNoC architecture [8]

Table 7: Parameters for modeling the 3D organizations of our evaluated PNoCs.

| Parameters | CLOS PNoC [38] | SWIFT PNoC [8] |
| --- | --- | --- |
| Network Size | 256 cores | 256 cores |
| Network Radix | 8 | 16 |
| Network Diameter | 1 | 1 |
| Bisection Bandwidth (Gb/s) | 56×$N_\lambda$×BR | 32×$N_\lambda$×BR |
| Traffic Model | Multi-Threaded PARSEC Workloads [15] | |
| Photonic Layer Frequency | 5 GHz [83] | |
| Processing Core x86 Frequency | 2.5 GHz [83] | |
| TSV Channel Configuration Per Photonic Tx-Rx Block | 8 TSV Bundles | |
| TSV Bundle Size and Layout | 2×2 TSVs per Bundle [85] | |
| TSV Speed | 21 Gb/s [85] | |
| Energy of a TSV Bundle | 6.7 pJ [85] | |

The following subsection discusses the simulation results and how the modulation schemes compare against one another, for the two considered PNoC architectures.

## 5.2 Results and Discussion

### 5.2.1 Packet Latency

Figs. 8(a) and 8(b) show the average latency for different variants of CLOS-PNoC, for the different applications from the PARSEC benchmark suite [15], with all results normalized to CLOS_OOK for both the datarate-BER balanced and BER-optimal cases. It can be observed that the EDAC and ODAC variants of 4-PAM modulation on CLOS outperform the rest of the variants, and when compared to the baseline CLOS_OOK, they achieve 68% and 55% better latency for the balanced datarate-BER case, and 62% and 34% better latency for the BER-optimal case, on average. The 4-PAM-SS variant of CLOS displays 1.2× higher latency than the baseline for BER-optimal designs. The packet latency we observe is an indicator of the combined effect of the $N_\lambda$ and the data rates achieved for the links of these CLOS variants. Having a higher $N_\lambda$ increases the number of concurrent bits transferred over the network, which in turn reduces the packet transfer latency. Similarly, having a higher bit-rate increases the number of bits transferred in the given time frame, which also results in a lower packet latency. In addition, Utilizing 4-PAM allows us to transmit 2× bits per cycle for the same $N_\lambda$, allowing the 4-PAM signaling based variants of CLOS to have better latency values compared to their OOK based variants.

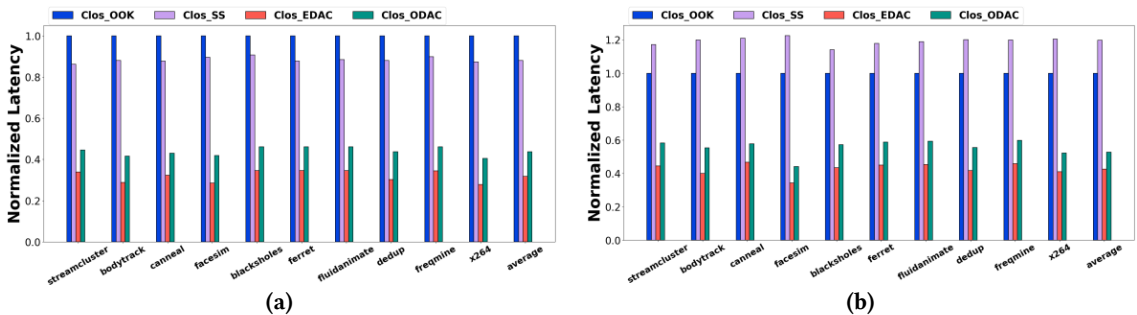

**Fig. 8:** Packet latency plotted across PARSEC benchmark applications [15] for (a) datarate-BER balanced variants, and (b) BER-optimal variants of CLOS PNoC. All results are normalized to the baseline CLOS_OOK.

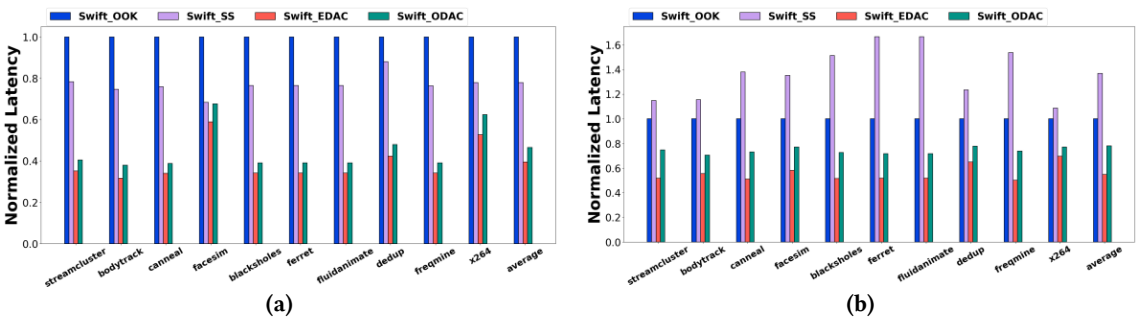

**Fig. 9:** Packet latency plotted across PARSEC benchmark applications [15] for (a) datarate-BER balanced variants, and (b) BER-optimal variants of SWIFT PNoC. All results are normalized to the baseline SWIFT_OOK.

Figs. 9(a) and 9(b) show the latency results for SWIFT PNoC, with results normalized to SWIFT-OOK, which acts as the baseline for our analysis. Here, again, the EDAC and ODAC 4-PAM variants obtain better latency values when compared to the baseline, due to the higher bandwidth they can achieve, as shown in Table 5. For these results, we can see that the EDAC variant performs better than other variants: 65% better latency than the baseline for the datarate-BER balanced case on average, as in 8(a), and 53% better latency on average for the BER-optimal case, as in 8(b). It can be noted that, similar to CLOS variants, for SWIFT PNoC as well, the SS variant has higher latency than the baseline for BER-optimal design, with 1.35× higher latency than the baseline on average. We can observe a similar trend in latency across the SWIFT PNoC variants, owing to the same reasons as discussed above, for the CLOS PNoC.

#### 5.2.2 Power Dissipation

Next, we examined the power dissipation in the considered PNoCs, with the results for CLOS PNoC shown in Fig. 10 and the results for SWIFT PNoC shown in Fig. 11. We report total power that is averaged across the considered PARSEC benchmark applications [15], and the corresponding error bars (with minimum and maximum values) are also shown in Figs. 10 and 11. The column heights in these figures show average values of total power, which is the sum of laser power (wall-plug power of laser sources), electrical power (power consumption of intra-cluster/intra-node communication in the electrical domain), TxRx power (dynamic power consumption of operating receiver/transmitted modulators, other devices, and the E/O and O/E conversion modules in PNoCs; per-MR values from Table 2), and total MR tuning power (sum of MR tuning power + microheater power; per-MR values from Table 2). The link-level results obtained for various CLOS and SWIFT links that are shown in Table 5 and Table 6 are directly reflected in the power results shown in Figs. 10 and 11. The wall-plug laser power values in Figs. 10 and 11 are directly dependent on the optical laser power values given in Tables 5 and 6. The higher optical laser values in Tables 5 and 6 translate into higher wall-plug laser power values in Figs. 10 and 11. Along the same lines, the TxRx and MR tuning power values in Figs. 10 and 11 depend on the $N_\lambda$ values from Tables 5 and 6; the higher $N_\lambda$ values translate into higher TxRx and MR tuning power values, as the number of MRs employed in a PNoC architecture depends on $N_\lambda$ and TxRx and MR tuning power depend on the number of MRs. Similarly, the higher values of $N_\lambda$ in Tables 5 and 6 also result in higher values of intra-cluster/intra-node electrical communication power in Figs. 10 and 11, as the sizes of the required electronic buffers in the intra-cluster electrical routers in PNoCs depend on the $N_\lambda$ values, and these buffer sizes in turn control the electrical power consumption in these routers.

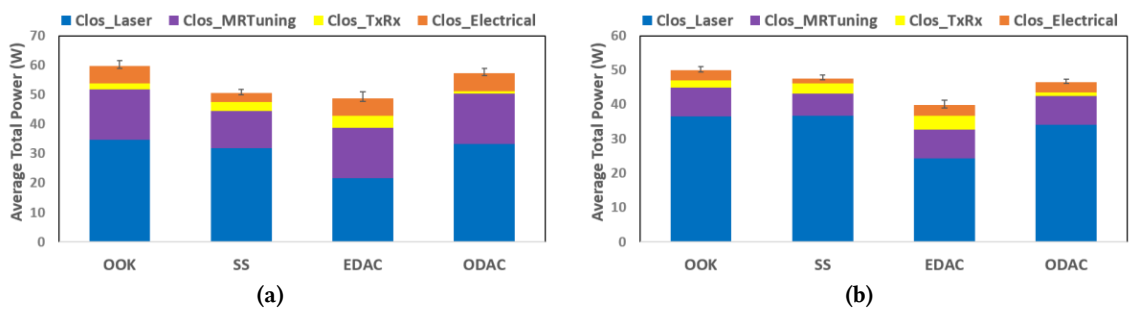

**Fig. 10:** Average total power dissipation for different (a) datarate-BER balanced, and (b) BER-optimal variants of CLOS PNoC. The error bars represent the minimum and maximum values of power dissipation across 12 PARSEC benchmarks [15].

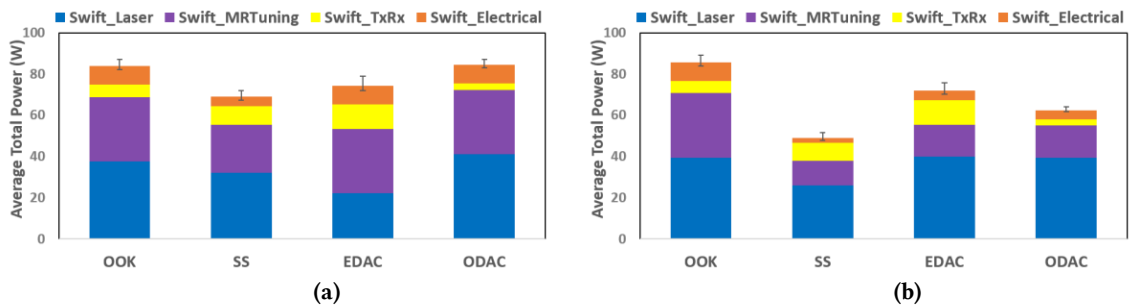

**Fig. 11:** Average total power dissipation for different (a) datarate-BER balanced, and (b) BER-optimal variants of SWIFT PNoC. The error bars represent the minimum and maximum values of power dissipation across 12 PARSEC benchmarks [15].

From Fig. 10, it can be observed that among the CLOS PNoC variants, the EDAC variants for both the datarate-BER balanced and BER-optimal cases dissipate the least power compared to other variants. This is because the laser power dissipation is the major contributor to the total power dissipation in CLOS PNoC, and the constituent links of the EDAC variants of the CLOS PNoC dissipate the lowest optical laser power (Tables 5 and 6) compared to the SS, ODAC, and OOK variants. To further analyze the power results, after the laser power, the second major contributor to the total power dissipation in CLOS PNoC is the MR tuning power, followed by the electrical power and TxRx power. The MR tuning power varies across different variants, because different variants require different number of MRs due to different $N_\lambda$ values. In contrast, the SS variants dissipate less electrical power compared to the other variants, because the SS variants achieve smaller $N_\lambda$ values, which in turn reduces the complexity of the routers and GIs in the SS variants. Also, the ODAC variants dissipate less TxRx power compared to the other variants, because the ODAC variants consume less dynamic energy in the modulator drivers (as can be inferred from the $E^{Mod}$ values in Table 2)

From Fig. 11, among the SWIFT variants, the SS variants for both the datarate-BER balanced and BER-optimal cases dissipate the least total power. This is because the constituent links of the SS variants achieve the smallest $N_\lambda$ (Tables 5 and 6), due to which these variants dissipate the lowest amount of electrical power, TxRx power, and laser power, the combined effect of which results in the lowest total power for these variants.

### 5.2.3 Energy-per-Bit

The energy-per-bit (EPB) results for CLOS variants are shown in Figs. 12(a) and 12(b), and for SWIFT PNoC in Figs. 13(a) and 13(b). Fig. 12(a) shows the EPB results for datarate-BER balanced variants of the CLOS PNoC, where EDAC and ODAC variants of the CLOS PNoC have better EPB values, on average across PARSEC benchmarks [15], in comparison with the other CLOS variants, with 15% and 11% lower EPB, respectively, than the OOK variant. This is because of the higher aggregate data rate and lower packet latencies of the EDAC and ODAC variants resulting in lower energy consumption. For the BER-optimal case, as shown in Fig. 12(b), the OOK variant for CLOS PNoC can be observed to have much better performance, than the ones utilizing 4-PAM techniques. Among the 4-PAM techniques, using SS has substantially higher energy utilization, with 4.9× more EPB than the baseline OOK on average. Both CLOS ODAC and CLOS EDAC variants exhibit ~1.8× EPB of the OOK baseline for the BER-optimal case.

Among different SWIFT PNoC variants for the datarate-BER balanced case (Fig. 13(a)), we can see that the EDAC variant performs better across the benchmark applications, retaining 2.5× less EPB than the baseline OOK, on average across the PARSEC benchmarks [15]. The ODAC variant has comparable EPB consumption to the EDAC variant, with both consuming ~2.4× less EPB than the baseline on average across the benchmark applications. The SS variant also performs better than the baseline, consuming 1.2× less EPB on average across the benchmarks.

Among the BER-optimal variants of the SWIFT PNoC (Fig. 13(b)), the SS variant has ~1.7× more EPB value than the baseline OOK variant. On the other hand, the EDAC variant consumes 2.22× less EPB than the baseline OOK on average across the benchmarks. The ODAC variant consumes 1.67× less EPB than the baseline on average. For the SWIFT PNoC variants, as well as for the CLOS PNoC variants, the energy utilized by the laser sources and TxRx modules is the main factor controlling the EPB values, as seen in the power breakdowns in Figs. 10 and 11. The reduced power consumption of EDAC variants as shown in Figs. 10 and 11 along with their lower latency of operation, leads to better throughput, and results in better EPB values for these variants for both the PNoCs considered in our evaluations.

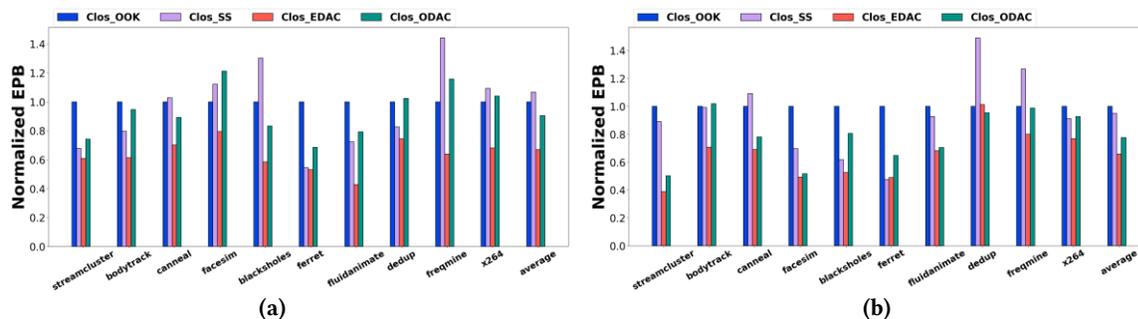

Fig. 12: Energy-per-bit (EPB) analysis for (a) the datarate-BER balanced variants, and (b) the BER-optimal variants of the CLOS PNoC. Column heights represent EPB averaged across 100 PV maps and normalized to the CLOS-OOK variant.

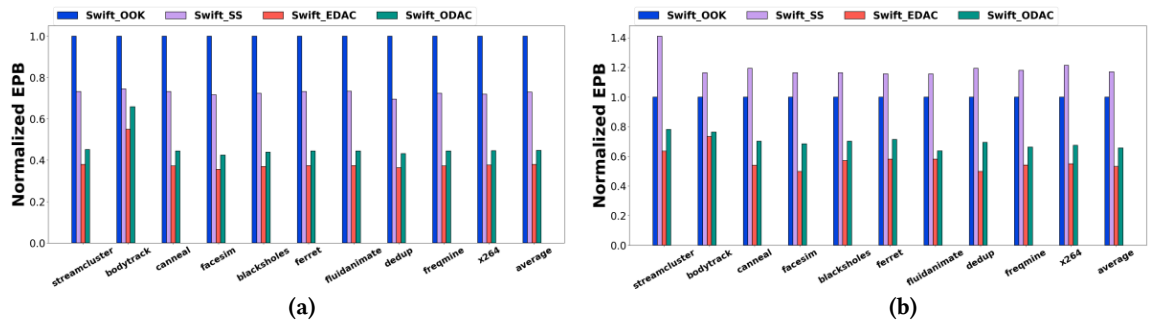

**Fig. 13: Energy-per-bit (EPB) analysis for (a) the datarate-BER balanced variants, and (b) the BER-optimal variants of the SWIFT PNoC. Column heights represent EPB averaged across 100 PV maps and normalized to the SWIFT-OOK variant.**

*In summary*, across all the OOK and 4-PAM variants of the CLOS and SWIFT PNoCs, the 4-PAM-EDAC variants exhibit lowest latency and energy on average across the considered PARSEC benchmark applications [15]. For the balanced datarate-BER case, compared to the baseline OOK variants, the 4-PAM-EDAC variants of the CLOS and SWIFT PNoCs achieve 68% and 65% of the latency, as well as 66% and 64% of the EPB. Similarly, for the optimal BER case, compared to the baseline OOK variants, the 4-PAM-EDAC variants of the CLOS and SWIFT PNoCs achieve 62% and 53% of the latency, as well as 38% and 57% of the EPB. These outcomes motivate the use of 4-PAM-EDAC signaling over OOK and other 4-PAM signaling methods to achieve significantly better energy-efficiency for on-chip communication with PNoCs.

# 6 Conclusion

Conventional OOK based signaling enables high-bandwidth parallel data transfer in PNoCs, but as the number of DWDM wavelengths increases, the power, area consumption, and bit-error rate (BER) in PNoCs increase as well. To address this problem, 4-PAM signaling has been introduced which can double the aggregated datarate without incurring significant area, power, and BER overheads. In this paper, for the first time, we performed a detailed analysis of various designs of 4-PAM modulators, including 4-PAM-SS, 4-PAM-EDAC, and 4-PAM-ODAC. We utilized these modulators to design 4-PAM photonic links and PNoC architectures with two different design goals of achieving the BER-balanced datarate (achieving maximum datarate with a desired BER of $10^{-9}$ using FEC codes) and optimal BER (achieving desired BER of $10^{-9}$ using increased input optical power). We then compared these BER-optimal and datarate-BER balanced 4-PAM links and PNoC architectures with the conventional OOK modulator based photonic links and architectures, in terms of performance (datarate and latency), BER, and energy-efficiency. Our analysis with CLOS PNoC and SWIFT PNoC architectures that are designed using the OOK, 4-PAM-SS, 4-PAM-EDAC and 4-PAM-ODAC based modulators and links showed that the 4-PAM-EDAC variants of the CLOS and SWIFT PNoCs yield the least latency and consume the least energy on average across the considered PARSEC benchmark applications [15]. For the balanced datarate-BER case, compared to the baseline OOK variants, the 4-PAM-EDAC variants of the CLOS and SWIFT PNoCs respectively achieve 68% and 65% of the baseline latency, as well as 66% and 64% of the baseline EPB. Similarly, for the optimal BER case, compared to the baseline OOK variants, the 4-PAM-EDAC variants of the CLOS and SWIFT PNoCs respectively achieve 62% and 53% of the baseline latency, as well as 38% and 57% of the baseline EPB. These outcomes push for the PNoC architectures of the future to employ the 4-PAM-EDAC signaling over the OOK and other 4-PAM signaling methods to achieve significantly better energy-efficiency for on-chip communication.

**Acknowledgment**

This research is supported by grants from NSF (CCF-1813370, CCF-2006788).